\documentclass[aps,prd,twocolumn,superscriptaddress,nofootinbib,10pt]{revtex4-2}

\usepackage{graphicx}
\usepackage{subcaption}        

\usepackage{textcase}
\usepackage{amsmath,amssymb,amsfonts}
\usepackage[a4paper, total={6.9in, 10in}]{geometry}
\usepackage{tcolorbox}
\usepackage{comment}
\usepackage{physics}
\usepackage{placeins}

\makeatletter
\renewcommand\section{\@startsection{section}{1}{\z@}%
  {-3.5ex \@plus -1ex \@minus -.2ex}%
  {2.3ex \@plus.2ex}%
  {\normalfont\bfseries\centering\protect\MakeTextUppercase}}
\renewcommand\subsection{\@startsection{subsection}{2}{\z@}%
  {-3.25ex\@plus -1ex \@minus -.2ex}%
  {1.5ex \@plus .2ex}%
  {\normalfont\bfseries}}
\makeatother

\usepackage{silence}
\begin{document}

\title{ Two-dimensional matter-wave interferometer, rotational dynamics, and spin contrast}

\author{Ryan Rizaldy}
\affiliation{
Van Swinderen Institute, University of Groningen, 9747 AG Groningen, The Netherlands\\}
\author{Shrestha Mishra}
\affiliation{Indian Association for the Cultivation of Science, Kolkata, India\\}
\author{Anupam Mazumdar}
\affiliation{
Van Swinderen Institute, University of Groningen, 9747 AG Groningen, The Netherlands\\}

\begin{abstract}
We investigate a two-dimensional matter-wave interferometer where both spatial and rotational dynamics of a nanoparticle are intertwined in closing the one-loop interferometer  in the Stern-Gerlach type setup. We consider the spin-contrast of the nitrogen-vacancy (NV) centred nanodiamond in combination with a two-dimensional magnetic field setup to extend the one-dimensional Stern--Gerlach interferometry. We analyse the dynamical motion along with the rigid rotation under the influence of the external magnetic field. Regarding rotation, we incorporate Euler-angle dynamics to analyse the stability of rotational degrees of freedom and their influence on the spin contrast to address the Humpty-Dumpty problem. We show that by imparting external rotation provides the gyroscopic stability to the liberating mode of the NV-spin and hence helps to improve the contrast. Our scheme creates a tiny spatial superposition of size $\sim 0.21~{\rm \mu m}$ for mass $m=10^{-17}$kg in less than $t\sim 0.013$s. 
\end{abstract}

\maketitle

\section{Introduction}

A massive spatial quantum superposition is becoming a well-sought experimental test of the foundations of quantum mechanics (for massive objects in quantum superposition)~\cite{bassireview} and of gravity. However, realising these experiments are filled with many theoretical and technical challenges~\cite{Bose:2025qns}. One of the prime reasons to create massive quantum superposition is to test the quantum nature of gravity/spacetime in a weak curvature imit in the lab~\cite{Bose:2017nin,ICTS}, the protocol is known as the QGEM (quantum gravity induced entanglement of matter) via witnessing the spin entanglement between the two massive nanodiamonds (embedded with a nitrogen vacancy (NV) center) in quantum superpositions kept adjacent to each other ~\cite{Bose:2017nin}, see also~\cite{Marletto:2017kzi}. The theoretical motivation for testing the quantum nature of gravity is very clear, the two quantum matter exchanges virtual graviton to entangle them through the quantum interaction between matter and massless spin-2 graviton
see~\cite{Marshman:2018upe,Bose:2022uxe,Vinckers:2023grv,Chakraborty:2023kel}. Extending this idea to an exchange of a massive graviton gives rise to new experimental bounds on the extra dimensions and the Kaluza-Klein graviton~\cite{elahi2023probing}, and 
the interaction between matter and photon via an exchange of a virtual graviton yeilds an entanglement between matter and photon, see~\cite{Biswas:2022qto}, which resembles the light-bending due to the curvature in the presence of a matter in general reklativity.

The core of the QGEM idea is to create a quantum spatial superposition  with the help of an internal spin of NV, by initiating a spin superposition in a motionally ground state of a nanodiamond. Nanodiamond being a rigid body, the rotational dynamics is equally important as highlighted in~\cite{ma2020quantum,Stickler18_GM,Rusconi:2022jhm, wachter2025gyroscopicallystabilizedquantumspin,ma2021torque,Rademacher:2025sye}. The motional and rotational coolings have been achieved in optical, ion and diamagnetic traps, see~\cite{Deli2020,Kamba:2023zoq,Piotrowski2023,Bykov:2022xji,Hsu:2016,schafer2021cooling}.
However, even if we cool the nanodiamond to a ground state and cool the rotational degrees of freedom, it is not sufficient to control the overall dynamics after the spin initialisation for the two arms of the matter-wave interferometer, because of the external torque on the NV spin experiences from the external magnetic field and its gradient in the Stern-Gerlach setup, see~\cite{Japha:2022phw,Japha2023}. The Stern-Geralch is a popular setup to create a macroscopic quantum spatial superposition, see~\cite{Wan16_GM,Scala13_GM,Pedernales:2020nmf,Marshman:2021wyk,Zhou:2022jug,Zhou:2022frl,Zhou:2024voj}.
The external torque leads to the mismatch in the Euler angles quantum wavepacket's overlap between the two arms, known as the Humpty-Dumpty problem~\cite{Schwinger,Englert,Scully} . Lack of spin contrast leads to a lack of visibility and affects the spin readout of NV spin while performing the one-loop interferometer, which is essential to the success of the QGEM experiment~\cite{Bose:2017nin}. One of the ways to mitigate this challenge is to impart a rotational stability before the onset of creating the superposition, as shown in these papers~\cite{Zhou:2024pdl,Rizaldy2025_RotationalStability,Rizaldy:2026cln}.

Besides the rotational issue, there are many other reasons why the spin contrast may suffer; interaction of the NV with the ambient impurities~\cite{Doherty_2013, Zhou:2025jki}, blackbody emission and absorption, collisional decoherence~\cite{Romero-Isart:2011yun,Hornberger_2012,vandeKamp:2020rqh,Schut:2023tce,Schut:2021svd,Tilly:2021qef,Rijavec:2020qxd,Schut:2023eux},  current/magnetic field fluctuations~\cite{Fragolino:2023agd,Moorthy:2025bpz,Moorthy:2025fnu}, and lastly the random external jitter~\cite{Toros:2020dbf,Wu:2024bzd}. All these effects can cause major dephasing and decoherence in the matter-wave system.

The current paper solely concentrates on a very specific
scenario of extending the earlier works where the spin contrast has been able to be improved by imparting rotation to the nanodiamond along the NV axis at the very beginning of creating teh spatial superposition in the Stern-Gerlach setup.  Here, we wish to extend that to two spatioal dimensions. This is beacuse, unless we create a trap which enforces the spatial motion of a nanodiamond to be in one-spatial diemsnion, say in $x$ direction, it is very hard to avoid the two dimensional spatial superposition. This has to do with the fact that two dimensional magnetic field is required to create a harmonic oscillator, and a tiny spatial superposition of order 
${\cal O}(10)$nm for nanodiamond of mass $m\sim 10^{-15}$kg, see~\cite{Rizaldy:2026cln,Xiang:2026kwd}. Here, we will study the intertwined dynamics between the two-diemnsional spatial superposition and the evolution of the Euler angles of the nanodiamond, especially the libration mode of the nanodiamond. We will show that by imparting the rotation along the NV axis provides the gyroscopic stability for the nanodiamond and helps us to obtain an improved spin visibility. This study is unique from the point of view that for the first time we study the rotational problem in the two spatial dimensional motion of a nanodiamond, and study the spin contrast. 

We first study the setup and study the center of mass dynamics of a nanodiamond in two spatial harnmonic oscillator.
In section III, we discuss the angular dynamics of the nanodianond in the context of two spatial dimensions. In section 1V, we discus the spin contrast and study the wavedunction of the libration mode and study the overlap. In section V, we discuss oour resuylts for small angle libration mode and study the flcutuations for all the Eauler angles. Finally, we conclude our paper.

\section{Set-Up and Center Of Mass Dynamics}

We consider a levitated nanodiamond hosting a single negatively charged nitrogen-vacancy (NV) center, confined within a three-dimensional harmonic trapping potential. 
The NV center provides an internal electronic spin degree of freedom that couples to external magnetic fields through its magnetic dipole moment. 
In our setup, the nanodiamond experiences both the trapping potential and an externally applied inhomogeneous magnetic field, which generates a position-dependent Zeeman interaction between the NV spin and the field. 
This magnetic field acts as the key mechanism that enables a Stern-Gerlach type spin-motion coupling~\cite{Pedernales:2020nmf,Marshman:2021wyk,Zhou:2022frl,Zhou:2022jug}, allowing spin-dependent forces to act on the centre of mass of the nanodiamond.

Let us imagine the externally applied magnetic field is described by
\begin{align}
    \mathbf{\hat{B}} = (B_0 + \eta x)\, \hat{e}_x + (\zeta y)\, \hat{e}_y,
\end{align}
where \(B_0\) is the uniform background field along the \(x\)-direction, and \(\eta\) and \(\zeta\) are the magnetic-field gradients along the \(x\) and \(y\) axes, respectively. This field configuration introduces a position-dependent Zeeman interaction between the NV spin and the magnetic field, giving rise to a spin-dependent potential that couples to the translational motion of the nanodiamond in two dimensions, see~\cite{japha2022role,Japha2023,Zhou:2024pdl,Rizaldy2025_RotationalStability,wachter2025gyroscopicallystabilizedquantumspin}. 
Such a setup extends the conventional one-dimensional Stern-Gerlach configuration by allowing spin-dependent forces to act simultaneously along both spatial directions.

The harmonic trap ensures small oscillations about the equilibrium position, which allows us to treat the nanodiamond’s motion semiclassically. 
In the following, we develop the coupled equations of motion governing both the center-of-mass and the rotational degrees of freedom of the nanodiamond under this two-dimensional magnetic field~\cite{Zhou:2024pdl,Rizaldy2025_RotationalStability,Rizaldy:2026cln}. However, as mentioned in the introduction, this paper will be the first to consider the effects of rotation on a two-dimensional superposition scheme within the SGI setup. From Maxwell's equation $\nabla \cdot \mathbf{B} = 0$, the field in Eq.~(2) must satisfy $\eta + \zeta = 0$, giving $\zeta = -\eta$. 

\subsection{Interferometric Sequence}
The protocol is to first cool the motion of the nanodiamond and also the rotational degrees of freedom, close to the ground state. It is possible to achieve that now in various setups~\cite{Deli2020,Kamba:2023zoq,Piotrowski2023,Bykov:2022xji,Hsu:2016,schafer2021cooling}. Then,
the NV center spin is prepared in an equal coherent superposition of its $m_s=\pm1$ states,
\begin{equation}
|\psi_s(0)\rangle = \frac{1}{\sqrt{2}}\left(|+1\rangle + |-1\rangle\right),
\end{equation}
which subsequently evolves under the spin-dependent magnetic forces. In contrast to pulsed multi–stage Stern–Gerlach protocols~\cite{Folman2018,Amit2019,doi:10.1126/sciadv.abg2879}, we consider here a minimal interferometric sequence: 
the inhomogeneous magnetic field is switched on at \(t = 0\) and kept constant until a final time \(t = t_{\mathrm{close}}\), at which the two spin–dependent trajectories are recombined.  
The superposition is created by preparing the NV center in a coherent spin state at \(t=0\), after which the two spin components experience opposite magnetic forces induced by the field gradients. 
This leads to a spatial separation of the corresponding wave packets along both \(x\) and \(y\) directions.  

At time \(t_{\mathrm{close}}\), the spin–dependent trajectories re–overlap due to the symmetric nature of the force profile and the harmonic confinement, and the spatial branches are recombined.
In this work, we focus on the dynamics during the interval \(0 \leq t \leq t_{\mathrm{close}}\), and assume that the switching of the magnetic field is fast compared to the mechanical time scales of the nanodiamond, but adiabatic with respect to the NV spin dynamics, so that no unwanted spin transitions are induced during the ramp.

\subsection{Hamiltonian}
Under these considerations, the system Hamiltonian takes the form~\cite{Pedernales:2020nmf,Marshman:2021wyk}
\begin{align}
    H = \frac{\mathbf{p}^2}{2m} + \sum_{j=1}^{3} \frac{L_j^2}{2I_j}
    - \frac{\chi_\rho m \mathbf{B^2} }{2\mu_0}
    + \boldsymbol{\mu}\cdot\mathbf{B}
    + H_{\text{ZFS}},
    \label{eq:fullH}
\end{align}
The quantities appearing in Eq.~(\ref{eq:fullH}) are defined as follows. 
Here $\mathbf{p}$ is the linear momentum of the nanodiamond, $L_j = \mathbf{L}\cdot\hat{n}_j$ are the components of the angular momentum along the body-fixed principal axes $\hat{n}_j$ with corresponding moments of inertia $I_j$, and $\chi_\rho$ is the mass magnetic susceptibility of diamond ($\chi_\rho = -6.2\times10^{-9}\,\mathrm{m^3/kg}$). 
The term $\boldsymbol{\mu} = g_e \mu_B \mathbf{S}$ denotes the magnetic moment associated with the NV electronic spin, and $H_{\text{ZFS}}$ corresponds to the zero-field splitting (ZFS) Hamiltonian of the spin-triplet ground state.

The zero-field splitting term in Eq.~(\ref{eq:fullH}) is given by, see~\cite{Doherty_2013}
\begin{align}
    H_{\text{ZFS}} = D\!\left[S_\|^2 - \frac{S(S+1)}{3}\right] + E(S_2^2 - S_3^2),
    \label{eq:zfs_rewrite}
\end{align}
where $D$ and $E$ are the axial and transverse ZFS parameters, respectively. 
The parameter $D/h \simeq 2.87~\mathrm{GHz}$ corresponds to the splitting between the $\ket{m_s = 0}$ and $\ket{m_s = \pm 1}$ sublevels of the ground-state spin triplet ($S=1$). 
The transverse term proportional to $E$ arises from strain or lattice asymmetry within the diamond crystal and typically satisfies $E \ll D$, with $E/h$ of order a few megahertz. 
When the strain contribution is negligible, the Hamiltonian reduces to the purely axial form governed by the parameter $D$.

Combining the Zeeman interaction with the intrinsic zero-field splitting (ZFS) of the NV center, the effective spin Hamiltonian can be written as, see~\cite{Zhou:2024pdl,Rizaldy2025_RotationalStability}
\begin{align}
    H_s = \boldsymbol{\mu} \cdot \mathbf{B} 
    + D\!\left[S_\|^2 - \frac{S(S+1)}{3}\right] 
    + E(S_2^2 - S_3^2),
    \label{eq:spinH_general}
\end{align}
where $\boldsymbol{\mu} = g_e\mu_B\mathbf{S}$ is the magnetic moment of the NV electronic spin, and $D$ and $E$ are the axial and transverse ZFS parameters, respectively. 
We consider the ground-state spin triplet with $S = 1$, and take the NV symmetry axis to be aligned along the unit vector $\hat{n}_s$. 
The spin operator is therefore projected along this direction as $S_\parallel = \mathbf{S} \cdot \hat{n}_s$. 

Expressed in the basis $\{\ket{m_s = +1}, \ket{0}, \ket{m_s = -1}\}$, the Hamiltonian in Eq.~(\ref{eq:spinH_general}) takes the matrix form~\cite{Rizaldy2025_RotationalStability,Rizaldy:2026cln}
\begin{align}
    H_s =
    \begin{pmatrix}
        \mu B_\parallel & \tfrac{\mu B_\perp}{\sqrt{2}} e^{-i\gamma} & E \\
        \tfrac{\mu B_\perp}{\sqrt{2}} e^{i\gamma} & -D & \tfrac{\mu B_\perp}{\sqrt{2}} e^{-i\gamma} \\
        E & \tfrac{\mu B_\perp}{\sqrt{2}} e^{i\gamma} & -\mu B_\parallel
    \end{pmatrix}
    + \frac{D}{3}\, \mathbb{I},
    \label{eq:spinH_matrix}
\end{align}
where $B_\parallel$ and $B_\perp$ denote the components of the magnetic field parallel and perpendicular to the NV axis, respectively, and $\gamma$ is the azimuthal angle of the magnetic field with respect to the NV frame. 
The additive term proportional to the identity matrix represents a uniform energy shift and does not affect the spin dynamics.

Using the Feshbach projection formalism, the spin Hamiltonian in Eq.~(5) can be reduced to the subspace spanned by the states $\{\ket{m_s = +1}, \ket{m_s = -1}\}$. 
This projection effectively eliminates the $\ket{0}$ state, which is energetically separated from the $\ket{\pm 1}$ manifold by the zero-field splitting $D$ and weakly coupled through the magnetic field. 
As a result, the dynamics can be described entirely within the two-level subspace corresponding to the $\ket{m_s = \pm 1}$ spin projections.The eigenenergies of the projected two-level spin Hamiltonian are given by~\cite{Rizaldy2025_RotationalStability,Rizaldy:2026cln}
\begin{align}
    V_s^{\pm} = \pm \sqrt{(\mu B_\parallel)^2 + \left|\epsilon(B_\perp)\right|^2}
    + \left(\frac{D}{3} + \frac{\mu^2 B_\perp^2}{2D}\right),
    \label{eq:Vs}
\end{align}
where
\begin{align}
    \epsilon(B_\perp) = E + \frac{\mu^2 B_\perp^2}{2D} e^{2i\gamma}.
    \label{eq:epsilon}
\end{align}
When the magnetic field is weaker than about $10^3$~G and changes slowly at the NV position, the spin follows the adiabatic evolution of the Hamiltonian. 
In this limit, the NV spin remains in one of its instantaneous eigenstates, either $\ket{0}$ or $\ket{\pm}$, where $\ket{\pm}$ are linear combinations of the Zeeman states $\ket{m_s = \pm 1}$.The spin Hamiltonian can then be expressed in the adiabatic basis as
\begin{align}
    H_{\text{spin}} = V_s^+ \ket{+}\bra{+} + V_s^- \ket{-}\bra{-},
    \label{eq:Hspin}
\end{align}
where $V_s^{\pm}$ are the energy eigenvalues given in Eq.~(\ref{eq:Vs}).

The transverse magnetic field component $B_{\perp}$ can be neglected since its contribution to the energy appears only at second order, 
$\Delta E_{\perp} \sim \mu^2 B_{\perp}^2 / (2D)$, 
which is strongly suppressed by the large zero-field splitting $D \approx 2.87~\mathrm{GHz}$. 
In contrast, the parallel component $B_{\parallel}$ shifts the spin levels linearly, 
$\Delta E_{\parallel} \sim \mu B_{\parallel}$, 
and therefore dominates the spin Hamiltonian in the experimentally accessible field range.
With these considerations, the energy eigenvalues reduce to~\cite{Rizaldy2025_RotationalStability,Rizaldy:2026cln}
\begin{align}
    V_s^{\pm} = \pm \sqrt{(\mu B_{\parallel})^2 + E^2} + \frac{D}{3}.
    \label{eq:Vs_simplified}
\end{align}

\subsection{Equations Of Motion of COM}
In addition to the spin-dependent potential $V_{s\pm}$, the total energy also includes the diamagnetic contribution arising from the term $V_{\text{dia}} = -\chi_\rho m \mathbf{B^2} / (2\mu_0)$ in the full Hamiltonian [see Eq.~(\ref{eq:fullH})]. 
The total potential energy of the nanodiamond is therefore
\begin{align}
    V_{\text{tot}} = V_{\text{dia}} + V_s^{\pm}.
\end{align}
The corresponding force acting on the center of mass can be obtained from the spatial gradient of the total potential as
\begin{align}
    \mathbf{F} = -\nabla V_{\text{tot}} = -\nabla V_{\text{dia}} - \nabla V_s^{\pm}.
    \label{eq:force_total}
\end{align}
and the equations of motion:-
\begin{align}
    &\ddot{x}(t) \pm \frac{\mu^{2} B_{\|} \cos \beta}{m\sqrt{\left(\mu B_{\|}\right)^{2}+\left|E\right|^{2}}} \cdot \eta + \frac{\chi_\rho}{2 \mu_0} \frac{\partial \mathbf{ B^2}}{\partial x}  = 0\\
    &\ddot{y}(t) \mp \frac{\mu^{2} B_{\|} \sin \beta}{m\sqrt{\left(\mu B_{\|}\right)^{2}+\left|E\right|^{2}}} \cdot \eta+ \frac{\chi_\rho}{2 \mu_0} \frac{\partial \mathbf{ B^2}}{\partial y} = 0
\end{align}
The laboratory-frame field components \(B_x,B_y\) project onto the NV quantization axis and its transverse plane as
\begin{align}
    B_{\parallel} &= B_x\cos\beta + B_y\sin\beta, \\
    B_{\perp} &= B_x\sin\beta - B_y\cos\beta,
\end{align}
where \(\beta\) is the angle between the NV axis \(\hat{n}_s\) (or the body-fixed axis \(\hat{n}_1\)) and the laboratory \(x\)-axis \(\hat{e}_x\). 
The transverse field \(B_{\perp}\) is resolved in the NV (co-moving) frame as
\begin{align}
    B_2 &= B_{\perp}\sin\gamma, \qquad 
    B_3 = B_{\perp}\cos\gamma,
\end{align}
with \(\gamma\) the azimuthal angle of the transverse component in the NV frame. 
Note that \(B_{\perp}=\sqrt{B_2^2+B_3^2}\) by construction.

\begin{figure}
    \centering
    \includegraphics[width=1\linewidth]{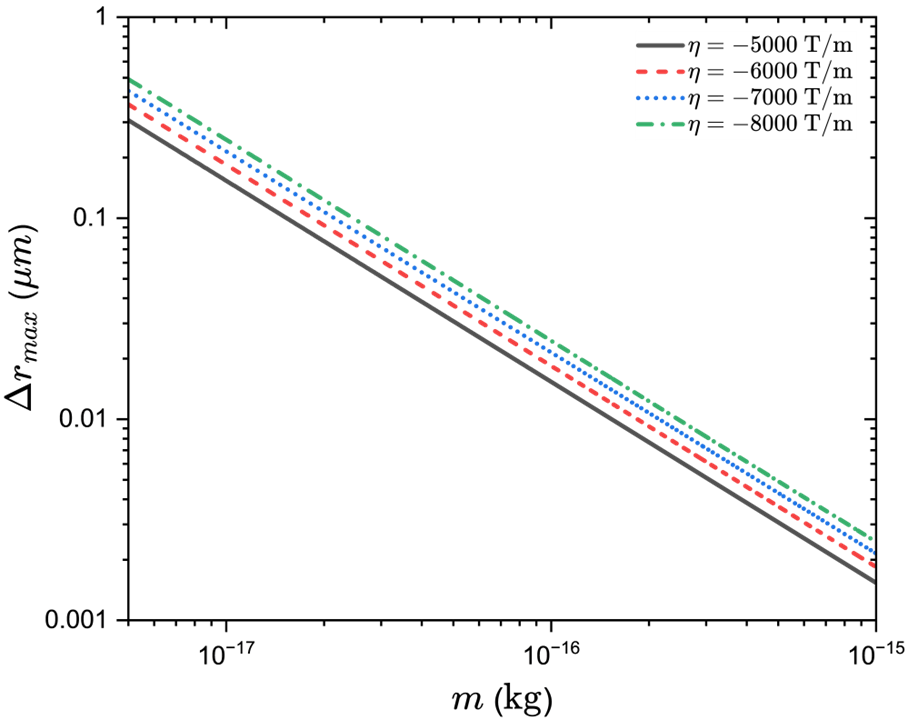}
    \caption{The maximum superposition size from Eq. \eqref{eq:superposition_max_size}, as a function of mass and variations in the magnetic field gradient, calculated using the approximation of libration mode $\beta \rightarrow \beta_0 = 10^{-3}$ rad and evaluated when the time reaches half a loop (at $t = \pi / \Omega$, with $\Omega = \sqrt{|\chi_\rho| \, \eta^2 / \mu_0}$).}
    \label{fig:maximum_superposition_size}
\end{figure}

For \(E \simeq 0\) the equations of motion reduce to
\begin{align}
\ddot{x}(t) + \frac{s\mu\eta}{m}\cos\beta + \frac{\chi_\rho\eta}{\mu_0}\,[B_0+\eta x(t)] &= 0, \label{eq:x_position}\\[4pt]
\ddot{y}(t) - \frac{s\mu\eta}{m}\sin\beta + \frac{\chi_\rho\eta^2}{\mu_0}\,y(t) &= 0 \label{eq:y_position},
\end{align}
which, using \(\Omega^2=\abs{\chi_\rho}\eta^2/\mu_0\) for the effective restoring coefficient, can be written as
\begin{align}
\ddot{x}(t) + \Omega^2 \left[x(t)+ \frac{B_0}{\eta}\right] + \frac{s\mu\eta}{m}\cos\beta &= 0, \label{eq:x_position2}\\[4pt]
\ddot{y}(t) + \Omega^2 y(t) - \frac{s\mu\eta}{m}\sin\beta &= 0 \label{eq:y_position2},
\end{align}
where \(s=\pm1\) labels the adiabatic spin branch and \(\mu\) is the NV magnetic moment. In fig.~\ref{fig:traj_beta}, we show a small spin-dependent splitting of the two trajectories, with maximum superposition size (see the derivation in Appendix \ref{appendix:superposition_size}) 
    \begin{align}
        \Delta r_{max}=\frac{4 \mu \eta}{m \Omega^2}\sqrt{1+\beta_0^2}, \label{eq:superposition_max_size}
    \end{align}
   where $\beta_0$ is the initial libration angle. In our case, $\Delta r_{max} \sim 0.215 \ \mu m$ at half loop time $t_{max} = \pi/\Omega$ for $m=10^{-17}$~kg (see Fig. \ref{fig:maximum_superposition_size} for general mass and gradient magnetic field cases).

\section{Angular Dynamics}

\subsection{Spherical nanorotor}

The nanodiamond is modeled as a rigid spherical rotor of mass \(m\) and radius \(R\), so that the principal moments of inertia are equal,
\[
I_1=I_2=I_3 \equiv I,
\]
and rotational dynamics are described by the rigid-body angular momentum \(\mathbf{L}\) and body-fixed axes \(\{\hat n_1,\hat n_2,\hat n_3\}\).

A right-handed laboratory frame \(\{\hat{e}_x, \hat{e}_y, \hat{e}_z\}\) is defined together with a body-fixed frame \(\{\hat{n}_1, \hat{n}_2, \hat{n}_3\}\) rigidly attached to the nanodiamond. The orientation of the body frame with respect to the laboratory frame is specified by the Euler angles \((\alpha, \beta, \gamma)\) following the XYX rotation sequence. 
In this convention, the angle \(\beta\) represents the rotation that aligns the laboratory axis \(\hat{e}_x\) with the body axis \(\hat{n}_1\).

\begin{figure}
    \centering
    \includegraphics[width=1\linewidth]{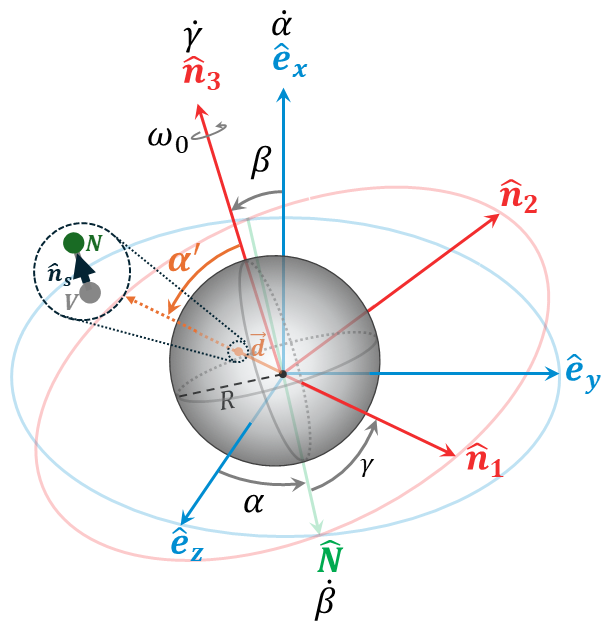}
    \caption{
    The illustration of the Euler-angle rotations with the fixed frame $\{x,y,z\}$ and the reference frame $\{\hat{n}_1,\hat{n}_2,\hat{n}_3\}$ of a spherical nanodiamond with radius $R$, inside which an off-center NV spin is located at a distance $d$ from the center and have spin orientation $\hat{n}_s$ forms an angle $\alpha'$ and parallel to $\hat{n}_3$-axis ($\hat{n}_s \parallel \hat{n}_3$ ). There are three Euler angles: the precession $\alpha$, the nutation $\beta$, and the rotation $\gamma$. An initial rotational velocity $\omega_0$ is applied around the $\hat{n}_3$-axis for the gyroscopic stability, see the discussion in section-C. We also set initial position of $\beta$ as $\beta(0)=\beta_0 = 10^{-3}$ rad.}
    \label{fig:euler_angle}
\end{figure}

In addition to specifying the rotational geometry, the position of the NV center within the nanorotor also needs to be fixed with respect to the center of mass. The NV electronic spin is assumed to be located at a point inside the nanorotor, displaced from the center of mass by a vector \(\vec{d}\) of magnitude \(d\). 
This displacement vector lies in the \(x\!-\!y\) plane of the laboratory frame and makes an angle \(\theta\) with respect to the \(\hat{e}_x\) axis, so that
\begin{equation}
\vec{d} = d(\sin\theta\,\hat{e}_y + \cos\theta\,\hat{e}_x).
\end{equation}

The internal symmetry axis of the NV center is not, in general, aligned with \(\vec{d}\) but is instead tilted by a fixed angle \(\alpha'\) relative to it. 
For simplicity, the spin quantization axis is chosen to coincide with the first body-fixed axis, \(\hat{n}_s \parallel \hat{n}_1\). 
The corresponding angle characterizing the orientation of the NV axis in the laboratory frame is then
$\beta = \theta - \alpha'.$

With this convention, the unit vector along the NV axis is given by
\begin{equation}
\hat{n}_s = \sin\beta\,\hat{e}_y + \cos\beta\,\hat{e}_x.
\end{equation}

This choice fixes both the spatial location of the NV center relative to the center of mass and the orientation of its spin quantization axis with respect to the laboratory frame, which will be used in the subsequent derivation of the coupled rotational and spin dynamics.

For later reference we use the standard ranges \(\alpha\in[0,2\pi),\ \beta\in[0,\pi],\ \gamma\in[0,2\pi)\). 
With this convention the projection of any laboratory vector onto the NV axis is \(\mathbf{v}\cdot\hat n_1\), and time dependence of \(\hat n_j(t)\) is determined by the Euler-angle dynamics to be derived in Appendix \ref{appendix:Effective_hamiltonian}.

The instantaneous rotation of the nanodiamond is described by the angular velocity vector \(\boldsymbol{\omega}\). 
In the adopted XYX Euler-angle convention (see Fig. \ref{fig:euler_angle}), it can be expressed as
\begin{equation}
    \boldsymbol{\omega} = \dot{\alpha}\,\hat{e}_{x} + \dot{\beta}\,\hat{N} + \dot{\gamma}\,\hat{n}_{1},
    \label{eq:omega_general}
\end{equation}
where \(\hat{N}\) is an auxiliary unit vector normal to both \(\hat{e}_{x}\) and \(\hat{n}_{1}\). 
Projecting these unit vectors onto the body-fixed axes \(\{\hat{n}_1, \hat{n}_2, \hat{n}_3\}\) gives
\begin{align}
    \hat{N} &= \cos\gamma\,\hat{n}_{2} + \sin\gamma\,\hat{n}_{3}, \label{eq:N_projection}\\
    \hat{e}_{x} &= \cos\beta\,\hat{n}_{1} 
    + \sin\beta\,\sin\gamma\,\hat{n}_{2}
    - \sin\beta\,\cos\gamma\,\hat{n}_{3}. \label{eq:ex_projection}
\end{align}

\subsection{Magnetic field at the NV location}

We now relate the magnetic field at the location of the NV center to the field defined at the center of mass (C.O.M.) of the nanodiamond.

The external magnetic field is taken in the form introduced earlier in the setup section, where the magnetic field at the C.O.M. is denoted by \( \mathbf{B}_{\rm COM} \).
The NV center is displaced from the C.O.M. by a fixed vector \( \mathbf{d} \) and its symmetry axis is taken to be aligned with the body axis \( \hat{\mathbf{n}}_1 \).

Using the definition of \( \mathbf{d} \) and the magnetic field profile at the center of mass, the magnetic field experienced by the NV center is then
\begin{equation}
B_{\rm NV}
= B_{\rm COM} + {\eta}\, d \cos(\beta + \alpha').
\end{equation}

Here \( {\eta} \) is the magnetic field gradient, and \( \beta \) is the libration angle of the nanorotor.

In the regime of weak field gradients (\({\eta} d \ll B_{\rm COM}\)), we approximate
\begin{equation}
B_{\rm NV} \simeq B_{\rm COM},
\end{equation}
and therefore treat the difference between the field at the NV location and at the C.O.M. as a small perturbative correction in our analysis.

\begin{figure*}[t]
    \centering
    \includegraphics[width=\linewidth]{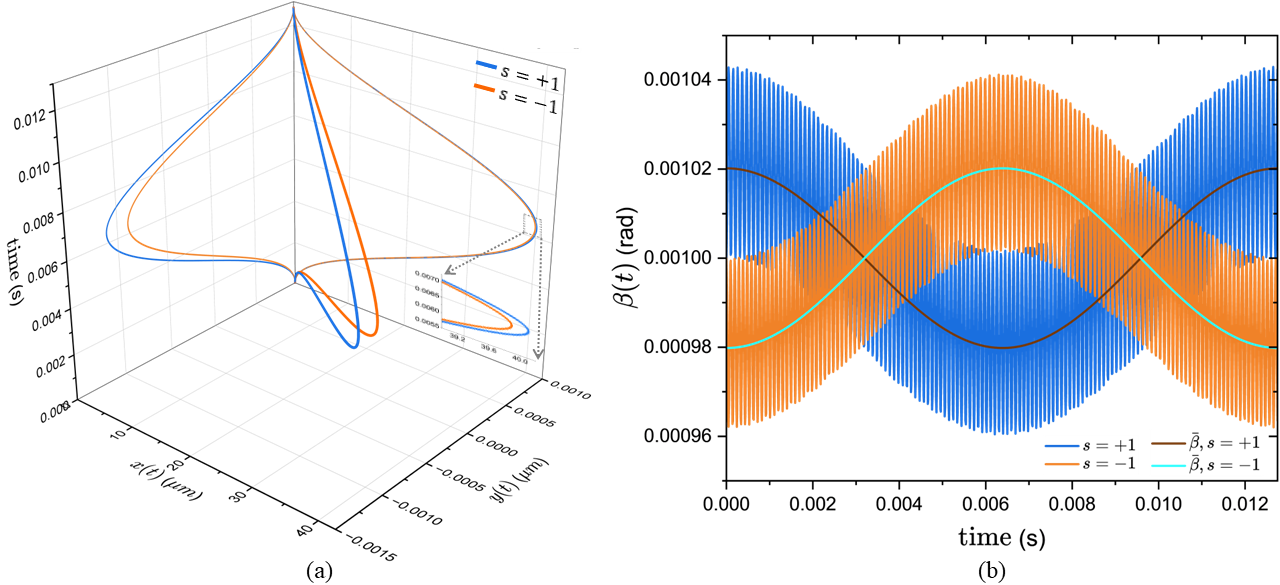}
    \caption{
    Numerical solutions of Eqs.~\eqref{eq:x_position}, \eqref{eq:y_position} and \eqref{eq:beta_eom_general} for a spherical nanodiamond with mass $m = 10^{-17}\,\mathrm{kg}$ and radius $R = 50\,\mathrm{nm}$.    
    For illustration, we take the center-of-mass motion is initialized at $x_0 = 0 \ \text{and} \ y_0 = 10^{-9}\,\mathrm{m}$ with $\dot x_0 = \dot y_0 = 0$, while the libration starts from $\beta(0)=\beta_0 = 10^{-3}\,\mathrm{rad}$ and $\dot\beta(0)=0$. In this plot we take the magnetic-field parameters $B_0 = 0.14\,\mathrm{T}$ and
    gradient $\eta = -7000\,\mathrm{T/m}$, and a libration frequency $\omega_0 = 2\pi\times 10^{4}\,\mathrm{s^{-1}}$. The evolution is shown up to $t_{\mathrm{close}} \simeq 0.01275\,\mathrm{s}$ 
    (corresponding to $t_{\mathrm{close}}=t_{close}=2\pi/\Omega$ with $\Omega^2 = 12.08\,\mathrm{s^{-2}}$
    ). Blue and orange curves correspond to the spin projections $s = +1$ and $s = -1$, respectively. (a) Three-dimensional spacetime trajectories $(x(t),y(t),t)$, showing a small spin-dependent splitting of the COM paths, with maximum superposition size $\Delta r_{max} = 0.215 \ \mu m$ at half loop time $t_{max} = \pi/\Omega$ (b) Time evolution of the libration angle $\beta(t)$ for the same parameters, illustrating that $\beta$ remains tightly confined around $\beta_0$ with only a small spin-dependent modulation, consistent with gyroscopic stabilization.}
    \label{fig:traj_beta}
\end{figure*}

\begin{figure*}
    \centering

    \begin{subfigure}[b]{0.47\linewidth}
        \centering
        \includegraphics[width=\linewidth]{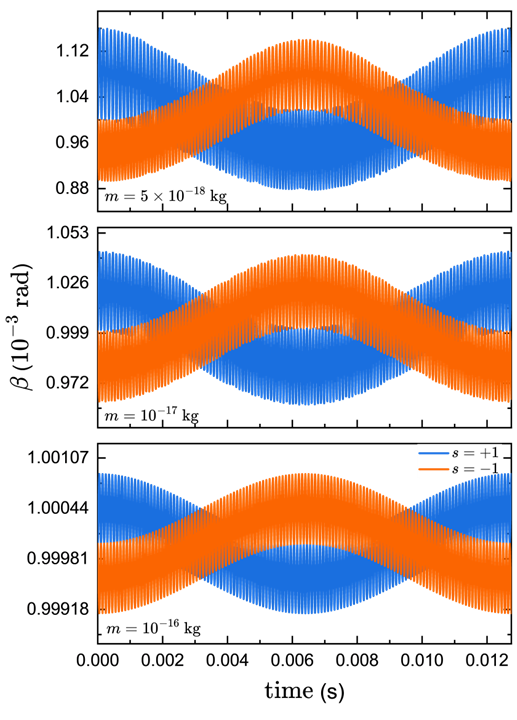}
        \caption{}
        \label{fig:beta_evols}
    \end{subfigure}
    \hfill
    \begin{subfigure}[b]{0.47\linewidth}
        \centering
        \includegraphics[width=\linewidth]{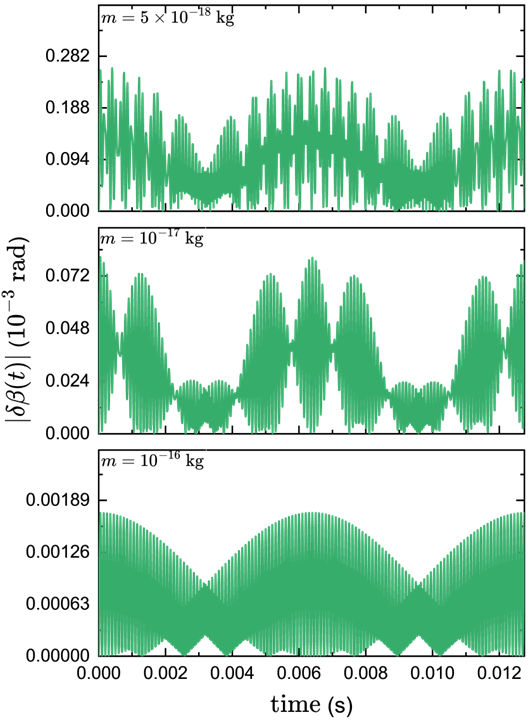}
        \caption{}
        \label{fig:beta_mismatch}
    \end{subfigure}

    \caption{
    (a) Time evolution of the libration angle $\beta(t)$ for spin projections $s=\pm1$ and three different nanodiamond masses $m=10^{-16},10^{-17},10^{-18}\,\mathrm{kg}$ (top to bottom). The trajectories are obtained by numerically solving Eqs.~(18), (19), and (40) with initial conditions $x_0 = y_0 = 10^{-9}\,\mathrm{m}$, $\dot x_0 = \dot y_0 = 0$, $\beta(0)=\beta_0 = 10^{-3}\,\mathrm{rad}$, and $\dot\beta(0)=0$, for a two–dimensional magnetic field characterized by $B_0 = 0.14\,\mathrm{T}$ and gradient $\eta = -7000\,\mathrm{T/m}$. The evolution is shown from $t=0$ to $t=t_{\mathrm{close}} = 2\pi/\Omega=0.01275$s, corresponding to two periods of the slow libration mode. For each mass, the spin states $s=\pm1$ produce slightly shifted mean values of $\beta(t)$ but the motion remains tightly confined around $\beta_0$, illustrating the strong gyroscopic stabilization of the NV axis.
    (b) Absolute angular mismatch $|\delta\beta(t)| = |\beta_{+}(t)-\beta_{-}(t)|$ for the same three masses and parameters as in Fig.~\ref{fig:traj_beta} (b). The mismatch remains well below a milliradian for $m=10^{-16}\,\mathrm{kg}$ and grows to a few milliradians as the mass is reduced to $10^{-18}\,\mathrm{kg}$, reflecting the $1/I$ scaling of the spin–torque–induced motion. Despite this increase, $|\delta\beta(t)|$ stays small over the whole interferometric time $0\le t\le t_{\mathrm{close}}$, indicating that the rotational dynamics of a fast-spinning, nearly spherical nanodiamond only weakly degrade the spin contrast in the regime considered.
    }
\end{figure*}

\begin{figure*}[t]
    \centering

    \begin{subfigure}[b]{0.45\linewidth}
        \centering
        \includegraphics[width=\linewidth]{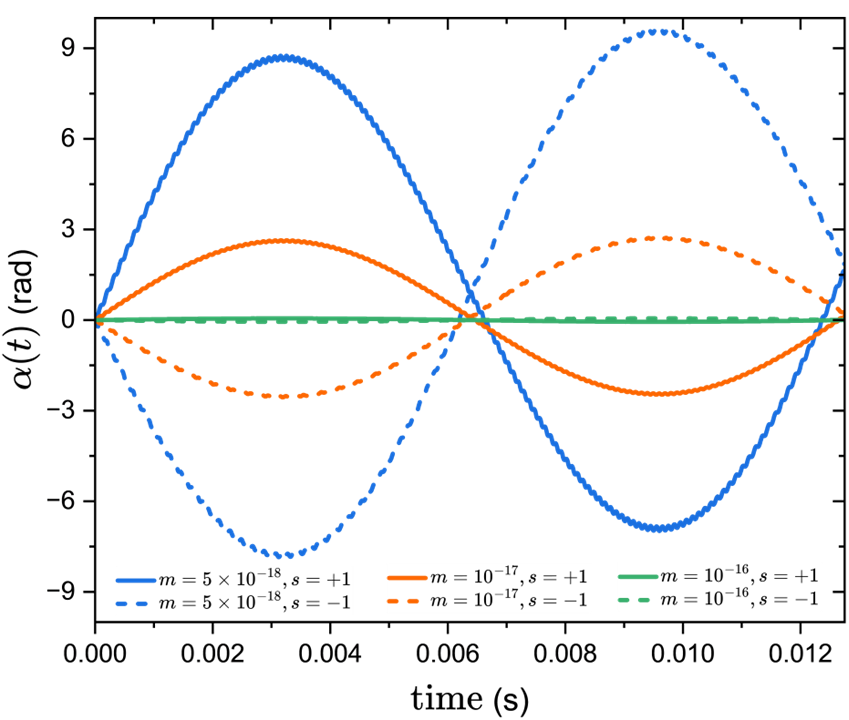}
        \caption{}
        \label{fig:alpha_t}
    \end{subfigure}
    \hfill
    \begin{subfigure}[b]{0.45\linewidth}
        \centering
        \includegraphics[width=\linewidth]{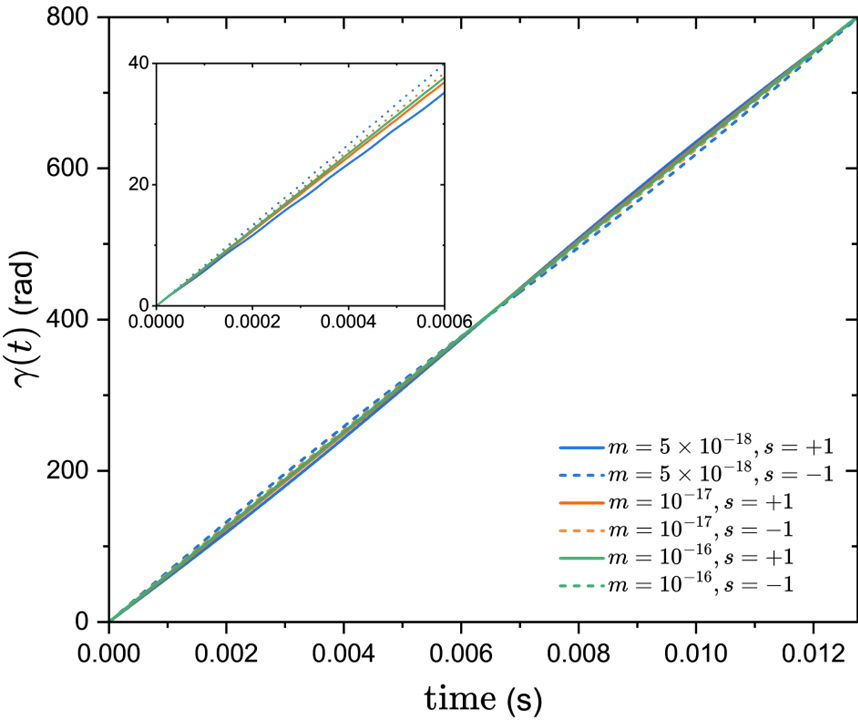}
        \caption{}
        \label{fig:gamma_t}
    \end{subfigure}
    \hfill
    \begin{subfigure}[b]{0.45\linewidth}
        \centering
        \includegraphics[width=\linewidth]{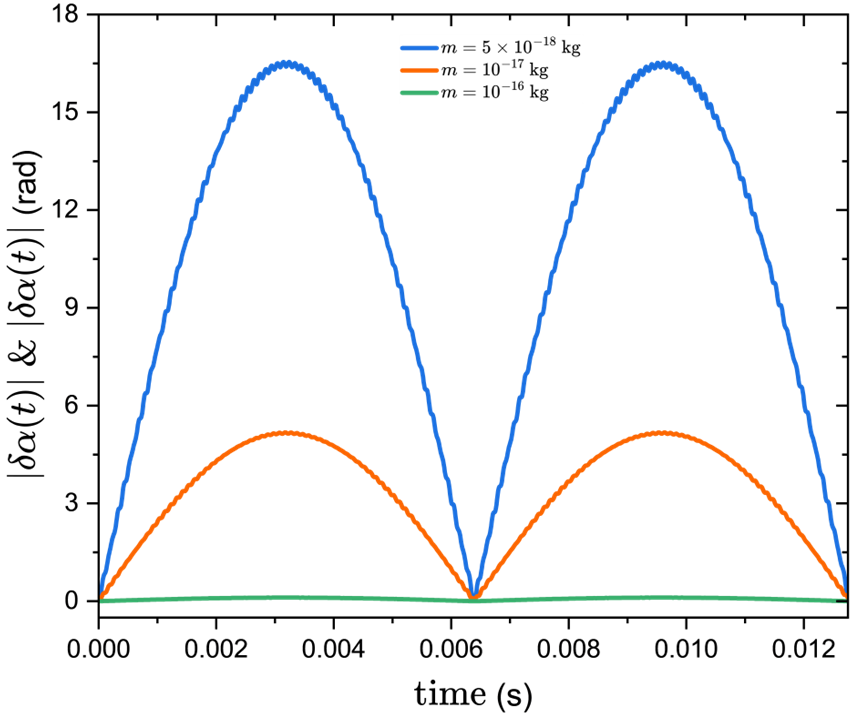}
        \caption{}
        \label{fig:alpha_gamma_mismatch}
    \end{subfigure}

    \caption{
    (a) Time evolution of the precession angle $\alpha(t)$ for spin states $s=\pm1$ and masses  $m=10^{-16},10^{-17},5\times10^{-18}\,\mathrm{kg}$, obtained by integrating Eq.~\eqref{eq:beta_eom_general} for $\beta(t)$ and inserting the result into Eq.~\eqref{Euler_angle-1}. Lighter particles show stronger spin-dependent precession due to their smaller rotational inertia. 
    (b) Time evolution of the rotation angle $\gamma(t)$ from Eq.~\eqref{euler_angle-2}, using the same  $\beta(t)$ as in panel (a). Note that at initial time $t=0$, $\gamma(0)=\omega_0$. The nearly linear growth reflects rigid-body rotation at frequency $\omega_0$, with a small spin-dependent splitting from the $\mu\cdot B$ contribution in the Hamiltonian. 
    (c) Precession and rotation-angle mismatch $\delta\alpha(t)=|\alpha_{+}(t)-\alpha_{-}(t)|$ and $\delta\gamma(t)=|\gamma_{+}(t)-\gamma_{-}(t)|$ for the same three masses. The mismatch increases for smaller masses, consistent with the scaling $\delta\alpha\propto I^{-1}$.
    Common parameters for all panels: $\beta(0)=\beta_0=10^{-3}\,\mathrm{rad}$, $\dot{\beta}(0)=0$, $\alpha(0)=0$, $\dot{\alpha}(0)=0$, $\gamma(0)=0$, $\dot{\gamma}(0)=\omega_0$, $x_0=0,y_0=10^{-9}\,\mathrm{m}$, $\dot{x}(0)=\dot{y}(0)=0$, $B_0=0.14\,\mathrm{T}$, $\eta=-7000\,\mathrm{T/m}$, and closure time $t_{\mathrm{close}}\simeq 2\pi/\Omega\,\mathrm{s}=0.01275$s.
    }
\end{figure*}

\subsection{Effective Spin Hamiltonian}

In the co-moving (body-fixed) frame, the spin degrees of freedom experience both the magnetic interaction and a coupling to the mechanical rotation of the nanodiamond. 
The effective spin Hamiltonian, including the Zeeman term and the Einstein--de Haas contribution, can be written as~\cite{Japha:2022phw,Japha2023,Zhou:2024pdl,Rizaldy2025_RotationalStability}
\begin{equation}
    \widetilde{H}_{s}
    = \mu\,\mathbf{S}\cdot\mathbf{B}
    + D\!\left[S_{\parallel}^{2} - \frac{S(S+1)}{3}\right]
    - \hbar\,\mathbf{S}\cdot\frac{\mathbf{L}}{\mathbf{I}},
    \label{eq:Heff_spin}
\end{equation}
where the first term describes the Zeeman interaction between the NV spin and the local magnetic field, the second term represents the intrinsic zero-field splitting (ZFS), and the third term accounts for the coupling between the spin and the rigid-body rotation. 

The additional term proportional to \(-\hbar\,\mathbf{S}\cdot\mathbf{L}/\mathbf{I}\) arises from the Einstein--de Haas effect~\cite{Stickler18_GM,Japha:2022phw,Japha2023}, which couples the spin angular momentum to the mechanical angular momentum of the nanodiamond. 
Physically, this term represents the rotational back-action produced by spin reorientation in a freely rotating body. 
Although typically small compared to the Zeeman and ZFS terms, their inclusion is necessary to capture spin-rotation dynamics in a levitated nanodiamond. 
A detailed derivation of this term is provided in Appendix~\ref{appendix:Effective_hamiltonian}.

With the initial conditions \(\dot{\gamma}(0)=\omega_{0}\) and \(\dot{\alpha}(0)=\dot{\beta}(0)=0\), and the initial angles 
\(\beta(0)=\beta_{0}\), \(\alpha(0)=\pi/2\), and \(\gamma(0)=0\), 
the Euler angles evolve as \(\alpha(t)=\pi/2\), \(\beta(t)=\beta_{0}\), and \(\gamma(t)=\omega_{0}t\). 
The magnetic field in the co-moving (body) frame is then
\begin{align}
\vec{B}_{\text{body}} =
\begin{bmatrix}
B_1\\ B_2\\ B_3
\end{bmatrix}
=
\begin{bmatrix}
B_x\cos\beta_0 + B_y\sin\beta_0 \\
B_x\sin\beta_0\sin\gamma - B_y\cos\beta_0\sin\gamma \\
- B_x\sin\beta_0\cos\gamma + B_y\cos\beta_0\cos\gamma
\end{bmatrix}.
\end{align}

Projecting the spin along \(\hat{n}_1\), the Zeeman and zero–field splitting terms combine with the Einstein–de Haas coupling to give a time–dependent effective spin Hamiltonian in the co–moving frame. 
The explicit matrix form and intermediate steps are derived in detail in Appendix~\ref{appendix:Effective_hamiltonian}. 
Here we only state the final result obtained after eliminating the \(|m_s=0\rangle\) manifold using the Feshbach projection formalism. 
Within the reduced two–level subspace \(\{|m_s=+1\rangle,\,|m_s=-1\rangle\}\), the effective NV Hamiltonian is
\begin{widetext}
    \begin{align}
    H^{\text{eff}}_{NV} &=
\begin{pmatrix}
(\mu B_{\|} - \hbar \omega_0) &
\dfrac{\mu^2 B_{\perp}^2 e^{-2i\gamma}}{2D}\\[3pt]
\dfrac{\mu^2 B_{\perp}^2 e^{2i\gamma}}{2D} &
-(\mu B_{\|} - \hbar \omega_0)
\end{pmatrix}+\left(\frac{D}{3}+\frac{\mu^2 B_{\perp}^2}{2D}\right) I 
\end{align}
\end{widetext}

The transverse magnetic component \(B_{\perp}\) is neglected for the same reasons discussed earlier: its contribution enters only at second order, 
\(\Delta E_{\perp}\sim \mu^2 B_{\perp}^2/(2D)\), and is strongly suppressed by the large zero–field splitting \(D\simeq2.87~\text{GHz}\). 
Provided the magnetic field strength remains well below \(10^3~\mathrm{G}\) and its spatial variation at the NV site is sufficiently slow, the spin dynamics is adiabatic. 
In this regime, the NV spin remains in one of the instantaneous eigenstates \(|+\rangle\) or \(|-\rangle\), which are superpositions of the Zeeman states \(|m_s=\pm1\rangle\).

The effective Hamiltonian in the adiabatic eigenbasis can therefore be written as
\begin{equation}
H_{\text{NV}}^{\text{eff}}
= V_{s+}\,|+\rangle\langle+| + V_{s-}\,|-\rangle\langle-| ,
\end{equation}
with the corresponding energy eigenvalues reducing to(for strain frequency, \(E \simeq 0\))
\begin{align}
    V_{s\pm} = \pm \mu B_{\parallel} + \frac{D}{3},
\end{align}
so that, in the adiabatic regime, the spin energies are determined solely by the longitudinal magnetic field component and the intrinsic zero–field splitting.

Moreover, to prevent nonadiabatic transitions such as Majorana spin flips~\cite{Zhou:2024pdl}, the rotational frequency \(\omega_0\) must remain much smaller than the spin precession frequencies, satisfying~\cite{einstein1915experimental,izumida2022einstein}
\begin{equation} \label{omega_constraints}
    \omega_0 \ll \frac{\mu B_{\|}}{\hbar}, 
    \qquad 
    \omega_0 \ll \frac{D \pm \mu B_{\|}}{\hbar}.
\end{equation}

\subsection{Evolution of the libration mode}

The total Hamiltonian of the spherical nanorotor is written as
\begin{equation}
    H_t = T_{\text{rot}} + H_{\text{NV}}^{\text{eff}},
\end{equation}
where \(T_{\text{rot}}\) describes the mechanical rotation of the rigid body and 
\(H_{\text{NV}}^{\text{eff}}\) accounts for the adiabatic spin contribution. 
The rotational kinetic energy in terms of the Euler angles is
\begin{equation}
    T_{\text{rot}} = \frac{I}{2}\!\left(\dot{\beta}^2 + \dot{\alpha}^2 \sin^2\beta\right)
    + \frac{I}{2}\!\left(\dot{\alpha}\cos\beta + \dot{\gamma}\right)^2.
\end{equation}
so that the total Hamiltonian becomes~\cite{Zhou:2024pdl,Rizaldy2025_RotationalStability,Rizaldy:2026cln}
\begin{align}
     H_t
    =& \ \frac{p_\beta^2}{2I}
    + \frac{p_\gamma^2}{2I}
    + \frac{(p_\alpha - p_\gamma\cos\beta)^2}{2I\sin^2\beta}\nonumber \\
    &+ s\mu(B_x\cos\beta + B_y\sin\beta).
    \label{eq:Ht_full}
\end{align}

Since the Lagrangian \(\mathcal{L}=T_{\text{rot}}-V\) depends only on the tilt angle \(\beta\), the coordinates \(\alpha\) and \(\gamma\) are cyclic.  
As a result, their conjugate momenta are conserved quantities of the motion. 
Evaluating them using the initial conditions \(\dot{\alpha}(0)=\dot{\beta}(0)=0\) and \(\dot{\gamma}(0)=\omega_0\), we obtain
\begin{align}
    p_\alpha &= I\omega_0\cos\beta_0, &
    p_\beta &= I\dot{\beta}, &
    p_\gamma &= I\omega_0.
\end{align}

The dynamics is therefore fully governed by the evolution of the single angle \(\beta(t)\). 
Hamilton’s equation for \(p_\beta\) gives
\begin{equation}
    \dot{p}_\beta
    = -\frac{\partial H_t}{\partial\beta}
    = -\frac{\partial}{\partial\beta}\!\left[\frac{(p_\alpha-p_\gamma\cos\beta)^2}{2I\sin^2\beta}\right]
    - \frac{\partial V_s}{\partial\beta}.
\end{equation}
Identifying \(p_\beta = I\dot{\beta}\), this leads to the second–order equation of motion
\begin{align}
\begin{split}
     I\ddot{\beta}
    =& \ \frac{I\omega_0^2(\cos\beta_0 - \cos\beta)(\cos\beta_0\cos\beta - 1)}{\sin^3\beta}\\
    &+ s\mu(B_x\sin\beta - B_y\cos\beta).
\label{eq:beta_eom_general}
\end{split}
\end{align}
where the first term represents the purely inertial contribution of the rigid-body rotation, 
while the second term accounts for the magnetic torque arising from the spin–field interaction.
This equation shows explicitly how the mechanical precession of the rotor is modified by the spin-dependent magnetic torque.

For completeness, the associated angular velocities of the remaining Euler angles are
\begin{align}
    \dot{\alpha} &= \omega_0\,\frac{\cos\beta_0 - \cos\beta}{\sin^2\beta}, \label{Euler_angle-1} \\[4pt]
    \dot{\gamma} &= \omega_0\!\left[1 - \frac{(\cos\beta_0 - \cos\beta)\cos\beta}{\sin^2\beta}\right],\label{euler_angle-2}
\end{align}
which describes the induced precession and spin of the nanorotor as it evolves under the combined action of inertia and magnetic coupling. At $t=0$, $\dot \gamma(0)=\omega_0$, also, $\beta(t=0)=\beta_0$, and $\alpha(t=0)=0$. Indeed, these are the initial conditions we are taking. Of course, changing them slightly will not affect our simulations, such as setting $\beta_0$ to a small libration angle and initially setting $\alpha=0$. During the evolution for creating spatial superposition and closing the superposition in two spatial dimensions, these Euler angles will evolve in time by Eqs.~(\ref{Euler_angle-1}),(\ref{euler_angle-2}), which we depict in Figs.\ref{fig:alpha_t},~\ref{fig:gamma_t}. In the next section, we will perform a small-angle analysis to examine how their evolution occurs in the SGI setup.

\subsection{Small-angle evolution of the libration mode}
We now analyze the rotational dynamics in the small–angle regime, where the tilt of the NV axis remains close to its initial value. 
Writing
\begin{equation}
    \beta(t) = \beta_0 + \delta\beta(t), 
    \qquad |\delta\beta| \ll \beta_0 \ll 1,
\end{equation}
we expand Eq.~\eqref{eq:beta_eom_general} to linear order in \(\delta\beta\).

To leading order, the equation of motion reduces to
\begin{equation}
\ddot{\beta}
= -\left(\omega_0^2 - \frac{s\mu}{I}\big[B_x + B_y\beta_0\big]\right)\delta\beta
+ \frac{s\mu}{I}\big(B_x\beta_0 - B_y\big),
\label{eq:libration_simplified}
\end{equation}
which has the structure of a harmonic oscillator with a spin–dependent equilibrium shift.  
It is convenient to rewrite this as
\begin{equation}
\ddot{\beta} 
= -\omega^2(t)\big[\beta - \bar{\beta}(t)\big],
\end{equation}
with the effective frequency and equilibrium position given by
\begin{align}
\omega^2(t) &= \omega_0^2 - \frac{s\mu}{I}\big[B_x(t) + B_y(t)\beta_0\big], \\[4pt]
\bar{\beta}(t) &= \beta_0 + 
\frac{s\mu\big[B_x(t)\beta_0 - B_y(t)\big]}{I\omega_0^2 - s\mu\big[B_x(t)+B_y(t)\beta_0\big]}.
\end{align}
Physically, \(\omega(t)\) represents a weak spin–dependent modulation of the natural rotational frequency, while \(\bar{\beta}(t)\) corresponds to a shift of the stable orientation due to the magnetic torque.  
Since the inertial term dominates in realistic parameters, \(\omega_0^2 \gg \mu B_x/I\), we may safely approximate
\begin{align}
\omega(t) &\simeq \omega_0,\\[4pt]
\bar{\beta}(t) &\simeq \beta_0 + \frac{s\mu}{I\omega_0^2}\big[B_x(t)\beta_0 - B_y(t)\big].
\end{align}

Under this approximation, the libration dynamics becomes
\begin{equation}\label{evolv_beta}
\beta(t) = A_\beta \cos(\omega_0 t) + \bar{\beta},
\end{equation}
with initial conditions \(\beta(0)=\beta_0\) and \(\dot{\beta}(0)=0\).  
The oscillation amplitude is therefore
\begin{equation}
A_\beta(t) = \frac{s\mu}{I\omega_0^2}\big[B_x(t)\beta_0 - B_y(t)\big],
\end{equation}
which measures the magnetic–torque–induced displacement from the unperturbed orientation.

In our interferometric protocol, the spin state is not flipped during the evolution. 
The two spin branches therefore experience opposite magnetic torques and follow slightly different rotational trajectories.  
The mismatch in their orientations at the recombination time \(t_{\text{close}}\) is given by
\begin{equation}
\delta\beta 
= \frac{4\mu}{I\omega_0^2}\big|B_x\beta_0 - B_y\big|.
\end{equation}

For the two–dimensional magnetic field profile
\begin{equation}
B_x(x) = B_0 + \eta x, \qquad B_y(y) = -\eta y,
\end{equation}
the initial mismatch becomes
\begin{equation} \label{mis-beta}
\delta\beta(0)
= \frac{4\mu}{I\omega_0^2}\big(B_0\beta_0 + \eta y_0\big),
\end{equation}
where \(y_0\) denotes the initial center-of-mass position of the nanodiamond along the \(y\)-direction, with \(x_0=0\) and \(y_0\neq 0\). 
This initial offset originates from the quantum zero-point fluctuations of the center-of-mass motion in the harmonic trap. 
The explicit derivation of \(y_0\) from the ground-state variance is given in Appendix~\ref{appendix:y_displacement}.

Since the spin remains adiabatically locked to its initial branch, the libration amplitude at recombination is unchanged:
\begin{equation}\label{miss-beta1}
A_\beta(t_{\text{close}}) = A_\beta(0).
\end{equation}
This shows that the spin–dependent magnetic torque induces a small oscillatory deviation of the NV axis, and that in the absence of spin flips the residual angular mismatch arises entirely from the different force profiles experienced by the two interferometer arms.
\subsection{Small angle evolution of $\alpha$ and $\gamma$}
For completeness, we also linearize the equations of motion for the precession angle $\alpha$ and the spin-rotation angle $\gamma$ around the small equilibrium angle $\beta_0$.  Using the relations obtained previously for a free rigid rotor and expanding to first order in $\delta\beta(t)=\beta(t)-\beta_0$ (with $\beta_0\ll1$), the evolution equations reduce to
\begin{align}
\dot{\alpha}(t) &\simeq \frac{ \omega_0}{\,\beta_0}\,\big[\beta(t)-\beta_0\big], \\
\dot{\gamma}(t) &\simeq -\,\frac{ \omega_0}{\,\beta_0}\,\big[\beta(t)-\beta_0\big] + \omega_0 .
\end{align}
The mismatch between the left and right interferometric trajectories in $\alpha$ and $\gamma$ is then obtained by integrating these equations with $\beta_{\pm}(t)$, yielding
\begin{equation}\label{alpha,gamma-diff}
\delta\alpha(t) = -\,\delta\gamma(t) 
\simeq \frac{\omega_0}{\beta_0}\int_0^{t}\!dt'\,\big[\beta_{+}(t')-\beta_{-}(t')\big].
\end{equation}
This last equation suggests that the difference in the angles $\ alpha$ and $ \gamma$ between the left and right trajectories of the SGI is exactly the same, which helps mitigate the Humpty-Dumpty problem in these angles. Hence, the main challenge remains mitigating the libration degree of freedom, as noted in Eqs.~(\ref{mis-beta}) and (\ref{miss-beta1}).

   \section{Numerical analysis of the superposition \& the libration mode including rotational stability}

In Fig.~\ref{fig:traj_beta}, we show the numerical solution of Eqs.~\eqref{eq:x_position}, \eqref{eq:y_position} for a spherical nanodiamond with mass $m = 10^{-17}\,\mathrm{kg}$ and radius $R = 50\,\mathrm{nm}$. We consider the center-of-mass motion is initialized at $x_0 = 0 \ \text{and} \ y_0 = 10^{-9}\,\mathrm{m}$. The $y_0\approx 10^{-9}$m is taken as a zero-point fluctuation along the $y$-direction. As mentioned earlier these values are for the purpose of illustration. We could have taken slightly different initial conditions for $x_0\sim 10^{-9}$m for $m\sim 10^{-17}$kg (corresponding to the Gaussian spread of the wavefunction in a harmonic trap), and the corresponding velocities $\dot x_0,~\dot y_0\sim 10^{-8}{\rm ms^{-1}}$, similar to that of the spread of the Gaussian wave packet for the above mass. However, this would not affect our main discussion, and it will not affect the plots to any significant precision at this point. It would definitely matter when we are designing a chip to perform the levitation and the superposition of our scheme, which we will perform in future publication.
In the dynamical equations, we also consider the initial libration d.o.f to be: $\beta(0)=\beta_0 = 10^{-3}\,\mathrm{rad}$ and $\dot\beta(0)=0$. We assume that we have significantly cooled the libration mode at the initiation. We take the magnetic-field parameters $B_0 = 0.14\,\mathrm{T}$ and the gradient $\eta = -7000\,\mathrm{T/m}$, and a libration frequency $\omega_0 = 2\pi\times 10^{4}\,\mathrm{s^{-1}}$. The evolution is shown up to $t_{\mathrm{close}} \simeq 0.01275\,\mathrm{s}$ 
    (corresponding to $t_{\mathrm{close}} =2\pi/\Omega$ with $\Omega^2 = 12.08\,\mathrm{s^{-2}}$). The blue and orange curves correspond to the spin projections $s = +1$ and $s = -1$, respectively. We see that the two-dimensional trajectory spans out in
    Three-dimensional spacetime $(x(t),y(t),t)$.  We also show the Time evolution of the libration angle $\beta(t)$ for the same parameters, illustrating that $\beta$ remains tightly confined around $\beta_0$ with only a small spin-dependent modulation, consistent with gyroscopic stabilization.

In Figs.~\ref{fig:beta_evols} and \ref{fig:beta_mismatch}, we show the evolution of $\beta(t)$ following Eq.~\ref{eq:beta_eom_general}, and its evolution for both the arms of the interferometer, i.e. $|\delta \beta(t)|=|\beta_{+}(t)-\beta_{-}(t)|$. One can see that by imparting the external rotation $\omega_0$ ( in the plots we have taken $\omega_0=2\times 10^{4}$ Hz helps them to stabilize the evolution of $\beta$. We work in the regime where we follow the constraints given by Eq.~(\ref{omega_constraints}). The large $\omega_0$ also dominates the second term in Eq.~(\ref{eq:beta_eom_general}), which can be seen in Eq.~(\ref{evolv_beta}), and the average value of $\beta(t) \approx \beta_0$. In Figs.~\ref{fig:beta_evols} and \ref{fig:beta_mismatch}, we also show the evolutions of $\beta(t)$ and $|\delta \beta(t)|=|\beta_{+}(t)-\beta_{-}(t)|$ for different masses, but the trend remains very similar. For heavy mass, both the amplitude of the oscillations are more suppressed by the additional effect of large moment of inertia, which is evident from varying the mass, i.e.,  $5\times10^{-18}- 10^{-16}$~kg.

In Figs. \ref{fig:alpha_t} and \ref{fig:gamma_t}, we show the evolution of two other Euler angles, $\alpha(t)$ and $\gamma(t)$, see Eqs.(\ref{Euler_angle-1}, \ref{euler_angle-2}). Since, by imparting initial rotation,  we are able to stabilize $\beta(t)\approx \beta_0$, one can see that $\dot \alpha \approx 0$, which leads to linear growth in $\alpha(t)$, which is evident from Fig.~\ref{fig:alpha_t}, while $\gamma(t)$ is slightly more susceptible to the slight difference between $\beta(t)$ and $\beta_0$, which is reflected in the sinus profile of small amplitude in Fig.~\ref{fig:gamma_t}. One can see again that this oscillation is highly suppressed for heavier mass, i,e. $m=10^{-16}$~kg as compared to $m=5\times 10^{-18}$~kg. Also, the difference between $\abs{\delta \alpha(t)} = \abs{\alpha_{+}(t)-\alpha_{-}(t)}= \abs{\delta \gamma(t)}= \abs{\gamma_{+}(t)-\gamma_{-}(t)}$, as we see in Eq.~(\ref{alpha,gamma-diff}). Fig.~\ref{fig:alpha_gamma_mismatch}, shows the evolution mismatch of the precessions and the rotation mode ($\abs{\delta\alpha}=\abs{\delta\gamma}$) at the recombination time given by $0.00168, 0.09058$, and $0.22073$ rad, for the nanodiamond masses, $m= 10^{-16},10^{-17},5\times 10^{-18}$ kg, respectively. 

Now, we discuss the spin contrast, and show how the external initial rotation helps to improve it via gyroscopic stability, see~\cite{Zhou:2024pdl,Rizaldy2025_RotationalStability}.

\section{Spin Contrast} 

\begin{figure}
    \centering
    \includegraphics[width=1\linewidth]{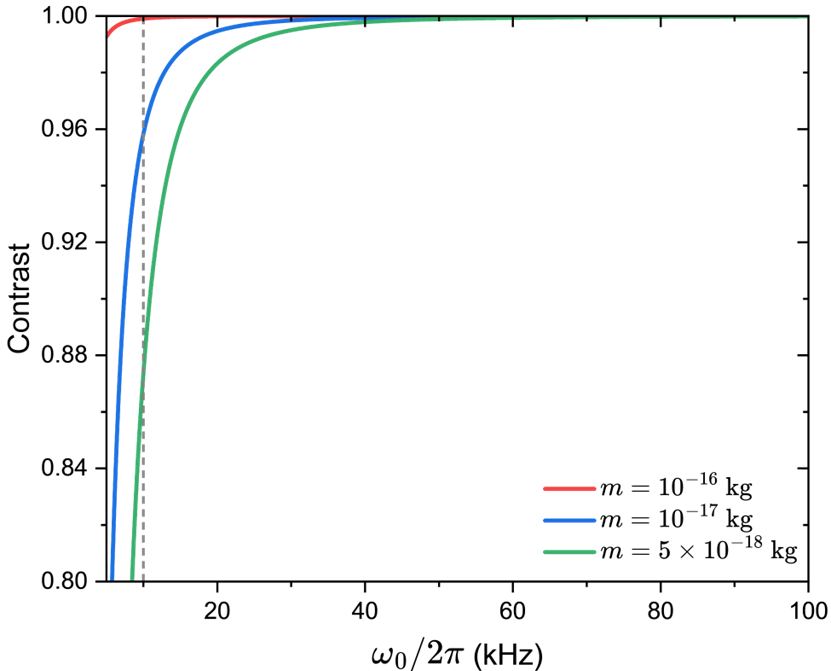}
    \caption{Plot from Eq. \eqref{eq:spin_contrast_final} which is the overlap of Gaussian wave packet or contrast value as a function of initial rotation ($\omega_0$), with $B_0 = 0.14$ T, $\eta=-7000$ T/m, $\beta_0 = 10^{-3}$ rad, and $y_0 = 10^{-9}$ m.  We also set the value of Gaussian width wave packet of precession and rotation modes as $\Delta p_\alpha=\Delta p_\gamma = 5\hbar$. We consider the masses, $m=\{10^{-16}, 10^{-17}, 5\times10^{-18}\}$ kg and shown the contrast. The values of the mismatch in ($\abs{\delta\alpha}=\abs{\delta\gamma}$) at the recombination time are 0.00168, 0.09058, and 0.22073 rad, for masses $m=\{10^{-16},~10^{-17},~5\times 10^{-18}\}$kg. The dashed gray line represent the set of initial rotation value ($\omega_0= 2\pi \times 10$ kHz) that we used in our SGI scheme.
   }
    \label{fig:contrast_mass}
\end{figure}

In the SGI setup, the measured signal is the population of the NV spin states at the output.  
The visibility of the interference fringes is therefore determined by the coherence between the two spin components of the SGI, which we quantify by the spin contrast between the left and right arms of the trajectories:
\begin{equation}
C = \big|\langle \Psi_L(t_{\text{close}}) | \Psi_R(t_{\text{close}}) \rangle \big|.
\end{equation}
Any imperfect recombination of the motional degrees of freedom (the ``Humpty–Dumpty'' effect) reduces this overlap and hence the contrast. Any imperfect recombination of the rotational degrees of freedom reduces the overlap and hence the contrast. Any small fluctuations lead to a loss in the contrast, and this is what we wish to minimise in a real experiment, see~\cite{Moorthy:2025bpz}. However, here we concentrate on the libration mode, while assuming that we are able to close the trajectories in such a way that the classical difference between the center of mass of the left and right trajectories, and their velocity differences are zero.

In our case, the relevant variables are the rotational degrees of freedom \((\alpha,\gamma,\beta)\) of the nanodiamond.  
Before the interferometric sequence is turned on, we assume that the angular wave packet is prepared as follows:
\begin{itemize}
    \item The angle of liberation \(\beta\) is in the ground state of a frequency harmonic trap \(\omega_0\), centered at the equilibrium angle \(\beta_0\).
    \item The angles \(\alpha\) and \(\gamma\) are described by Gaussian momentum distributions with widths \(\Delta p_\alpha\) and \(\Delta p_\gamma\) around their classical mean values.
\end{itemize}
The corresponding initial state can be written as, see~\cite{Zhou:2024pdl,Rizaldy2025_RotationalStability}
\begin{equation}
|\Psi(0)\rangle =
\int \frac{dp_\alpha' dp_\gamma'}{\sqrt{2\pi \Delta p_\alpha \Delta p_\gamma}}\,
e^{-\frac{p_\alpha'^2}{4\Delta p_\alpha^2}}
e^{-\frac{p_\gamma'^2}{4\Delta p_\gamma^2}}\,
|p_\alpha',p_\gamma'\rangle
\otimes |0\rangle_{\beta_0},
\end{equation}
where \(|0\rangle_{\beta_0}\) is the ground state of the libration mode.

During the interferometer, the magnetic-field gradient exerts opposite torques on the two spin components \(s=\pm 1\).  
This produces (i) slightly different classical rotational trajectories in \(\alpha\) and \(\gamma\), and (ii) spin-dependent displacements of the libration potential, so that the \(\beta\) mode evolves into two distinct coherent states.  
The final spin contrast is then obtained by propagating the above initial wave packet under the spin-dependent Hamiltonian and taking the overlap of the two resulting motional states.

The detailed derivation, including the linearization about the libration equilibrium and the coherent-state description of the \(\beta\) mode, is given in Appendix~\ref{appendix:spin contrast fromlibration}.  
Here we only quote the final result.  The contrast factorizes into a rotational contribution and a librational contribution, and we obtain the lower bound~\footnote{Note that this expression Eq.~\ref{eq:spin_contrast_final} is slight different from what we had obtained before in Refs.~\cite{Zhou:2024pdl,Rizaldy2025_RotationalStability,Rizaldy:2026cln}. First of all, here we do not take finite temperature correction, but the main difference is the presence of the last term, and in particular the presence of $\eta y_0$ contribution. This contribution is new and shows that the spin contrast explicitly depends on the initial condition for $y_0$ in the two-dimensional SGI setup.}
\begin{widetext}
    \begin{equation}
C \;\gtrsim\;
\exp\!\left[
-\frac{1}{2}
\left(
\delta\alpha^2 \frac{\Delta p_\alpha^2}{\hbar^2}
+ \delta\gamma^2 \frac{\Delta p_\gamma^2}{\hbar^2}
+ \frac{8\mu^2 (B_0\beta_0 + \eta y_0)^2}{I\hbar \omega_0^3}
\right)
\right].
\label{eq:spin_contrast_final}
\end{equation}
\end{widetext}

Here, \(\delta\alpha\) and \(\delta\gamma\) are the residual angular mismatches between the two spin paths at the closing time \(t_{\text{close}}\), while \(\Delta p_\alpha\) and \(\Delta p_\gamma\) are the corresponding momentum uncertainties of the initial wave packet.  
The last term originates from the separation of the two spin-dependent libration coherent states, where \(\omega_0\) is the libration frequency, \(I\) the moment of inertia, \(\mu\) the NV magnetic moment, \(B_0\) the homogeneous field, \(\eta\) the field gradient, and \(y_0\) the initial transverse position (given by the zero-point fluctuations of the center-of-mass mode; see Appendix~Y).
We note that the transverse displacement enters the contrast only through the
combination $(B_0\beta_0 + \eta y_0)$ in the last term of the exponent.  
The corresponding contribution to the contrast reads
\[
\exp\!\left[-\,\frac{4\mu^2 (B_0\beta_0 + \eta y_0)^2}{I\hbar\omega_0^{3}}\right],
\]
so the effect of the libration mode on $C$ is negligible whenever
\begin{equation}
\frac{4\mu^2 (B_0\beta_0 + \eta y_0)^2}{I\hbar\omega_0^{3}} \ll 1
\quad\Rightarrow\quad
|B_0\beta_0 + \eta y_0|
\ll
\frac{\sqrt{I\hbar\omega_0^{3}}}{2\mu}.
\label{eq:y0_condition}
\end{equation}
In practice, $y_0$ is set by the zero-point fluctuations of the
center-of-mass mode in the trapping potential (see Appendix~Y).  
Equation~\eqref{eq:y0_condition} therefore provides a constraint on the
allowed size of these fluctuations: for fixed $(B_0,\eta,\beta_0)$, the trap
should be tight enough (small $y_0$) or the rotation frequency $\omega_0$ large
enough that the inequality is satisfied.  

Eq.~\eqref{eq:spin_contrast_final} shows that all contributions to contrast loss are suppressed by large rotational frequency \(\omega_0\), illustrating the gyroscopic protection of spin coherence in a fast-spinning nanodiamond.
Fig.~\ref{fig:contrast_mass} shows the contrast in the presence of the initial rotation, i.e. $\dot \gamma(0)=\omega_0$ for different masses. Note that the contrast is the best for heavier masses, since the moment of inertia along with $\omega_0$ enter in the denominator of 
Eq.~(\ref{eq:spin_contrast_final}), higher is the combination, smaller is the exponent and higher becomes the contrast. Such a rotation of a nanoparticle (in the vicinity of $\sim {\cal O}(10)~{\rm KHz}$) is technologically feasible. In \cite{jin2024quantum}, the authors successfully levitated a nanodiamond containing an NV center in a high vacuum ($p<10^{-5}$ Torr) using a surface ion trap and rotated it up to 20 MHz. They observed the Berry phase induced by the mechanical rotation on the NV spin via ODMR, and demonstrated that rapid rotation can preserve spin coherence.

\section{Conclusion}

In this paper, we have calculated the SGI scheme using a simple linear external magnetic field, without any changes to the magnetic field gradient or spin flip operation as in Ref. \cite{Marshman2022,Zhou:2024pdl,Rizaldy2025_RotationalStability}.  We analyse a single-loop configuration of SGI with an inclusion of two spatial dimensions, which is necessary to accommodate Maxwell's equations and the presence of initial rotation to gyroscopically stabilise the NV-centered nanodiamond. We followed  \cite{Zhou:2024pdl,Rizaldy2025_RotationalStability} formalism to calculate the evolution of the libration modes of both the interferometric arms. 
This is the first such study of SGI in two spatial dimensions with an inclsuion of rotation, and we learned that it is possible to control both the spatial and the rotational dynamics to obtain a good contrast under certain conditions.

From the numerical calculations on the two-dimensional equations of motion, the two interferometric arms can reach a maximum superposiotion distance of about $0.215~\mu\text{m}$ for a nanodiamond mass of $10^{-17}~\text{kg}$. The two arms then recombine well, as indicated by $\delta x(t_{\text{closed}})=0$ and $\delta V_x(t_{\text{closed}})=0~\text{m/s}$. This indicates that the contrast loss due to the spatial-dyanmics can be minimised.

For the Euler angle dynamics, we take an initial libration angle of $\beta_0\sim 10^{-3}~\text{rad}$ and a fast initial rotational speed of $\omega_0 = 10~\text{kHz}$ (rotation along the NV axis) to produce gyroscopic stabilisation of the nanodiamond.
Throughout this setup, we are tacitly assuming that the coherence of the NV spin can be mintained. For the choice of the parameter, we expect the spin coherence to be maintained for roughly $t_{close}\sim 0.01275 $s.

The behavior of the $\beta$ angular evolution for a simple linear magnetic field is shown in Fig. \ref{fig:traj_beta}a, where both interferometric arms oscillate around nearly the same equilibrium angle $\bar{\beta}$. However, there is a phase difference due to the spin-state-dependent Zeeman effect, as seen from the $\delta\beta$ mismatch in Fig. \ref{fig:traj_beta}b. We also investigated the variation of the nanodiamond mass (assumed spherical) and obtained results consistent with Ref. \cite{Rizaldy2025_RotationalStability}: the larger the nanodiamond mass, the more stable it's gyroscopic behavior. This is evident from the smaller fluctuations in the $\beta$ angular mismatch for larger masses. This trend also holds for the rotation and precession angular mismatch, with $|\delta\alpha| = |\delta\gamma| = \{0.00168,\ 0.09058,\ 0.22073\}~\text{rad}$, which increases with decreasing nanodiamond mass. Keeping the mismatch of these three angles to a minimum is crucial to maintaining high contrast. To overcome the loss of contrast in the small mass of nanodiamond requires a faster initial rotation required, as shown in Fig. \ref{fig:contrast_mass}).

In the future, we will be considering a chip design which will mimic these features of SGI for a two dimensional spatial superposition in a diamagentically levitating setup. Hence, these very preliminary results will be extremely helpful for the critical design of the chip and realising them in an experimental setup step by step.

\begin{acknowledgments}
We would like to thank Sougato Bose for helpful discussions.
RR is supported by Beasiswa Indonesia Bangkit and Lembaga Pengelolah Dana Pendidikan (BIB LPDP) of the Ministry of Religious Affairs of Indonesia. A.M.'s research is funded by the Gordon and Betty Moore Foundation through Grant GBMF12328, DOI 10.37807/GBMF12328. This material is based on work supported by the Alfred P. Sloan Foundation under Grant No. G-2023-21130

\end{acknowledgments}

\bibliographystyle{apsrev4-2}
\bibliography{reference.bib} 

\clearpage
\appendix
\section{Effective Spin Hamiltonian}
\label{appendix:Effective_hamiltonian}
To describe the rotational dynamics of the system, the angular velocity can be written in terms of the time derivatives of the Euler angles. The total angular velocity takes the form
\begin{equation}
\boldsymbol{\omega}=\dot{\alpha} \hat{e}_{x}+\dot{\beta} \hat{N}+\dot{\gamma} \hat{n}_{1}, 
\end{equation}
where, $\hat{N}$ is the unit auxiliary vector. We now project each vector onto the $\mathbf{body -fixed } $ $\mathbf{axis}$:-

\begin{align}
\hat{N} & =\cos \gamma \hat{n}_{2}+ \sin \gamma \hat{n}_{3}  \\
\hat{e}_{x} & = \cos \beta \hat{n}_1 +\sin \beta \sin 
\gamma \hat{n}_2-\sin\beta \cos \gamma\hat{n}_{3}
\end{align}
 Then, the angular velocity of the comoving frame is :-
 \begin{equation}
     \boldsymbol{\omega}=\left[\begin{array}{c}
\omega_{1} \\
\omega_{2} \\
\omega_{3}
\end{array}\right]=\left[\begin{array}{c}
\dot{\alpha} \cos \beta+\dot{\gamma} \\
\dot{\alpha} \sin \beta \sin \gamma+\dot{\beta} \cos \gamma \\
-\dot{\alpha} \sin \beta \cos \gamma+\dot{\beta} \sin \gamma
\end{array}\right],
 \end{equation}
 and in the lab frame is:-
 \begin{equation}
     \boldsymbol{\omega}=\left[\begin{array}{c}
\omega_{x}\\
\omega_{y} \\
\omega_{z}
\end{array}\right]=\left[\begin{array}{c}
\dot{\gamma} \sin \beta \sin \alpha+\dot{\beta} \cos \alpha \\
\dot{\gamma}-\sin \beta \cos \alpha+\dot{\beta} \sin \alpha \\
\dot{\gamma} \cos \beta+\dot{\alpha}
\end{array}\right],
 \end{equation}
 and the transformation between the to frames:-
 \begin{widetext}
      \begin{equation}
     \left[\begin{array}{l}
\hat{n}_{1}\\
\hat{n}_{2} \\
\hat{n}_{3}
\end{array}\right]=\left[\begin{array}{ccc}
 \cos \beta& \sin \beta \sin \alpha & \sin \beta \cos \alpha\\
 \sin \beta \sin \gamma & \cos \alpha \cos \gamma-\cos \beta \sin \alpha \sin \gamma & -\sin \alpha \cos \gamma-\cos \beta \cos \alpha \sin \gamma\\
-\cos \gamma \sin \beta & \cos \alpha \sin \gamma+\cos \beta \sin \alpha \cos \gamma &  -\sin \alpha \sin \gamma+\cos \beta \cos \alpha \cos\gamma
\end{array}\right]\left[\begin{array}{l}
\hat{e}_{x} \\
\hat{e}_{y} \\
\hat{e}_{z}
\end{array}\right]
 \end{equation}
 \end{widetext}

Defining the effective spin Hamiltonian with the Zeeman term ($\mu(\mathbf{S} \cdot \mathbf{B})$) and the Einstein de - haas term, ($\hbar \mathbf{S} \cdot \frac{\mathbf{L}}{\mathbf{I}}$),
\begin{equation}
\widetilde{H}_{s}=\mu \mathbf{S} \cdot \mathbf{B}+D\left[S_{\|}^{2}-\frac{S(S+1)}{3}\right]-\hbar \mathbf{S} \cdot \frac{\mathbf{L}}{\mathbf{I}}
\end{equation}
written in the basis defined by the comoving frame. $A_{\|}$ is the component of a vector $\mathbf{A}$ parallel wrt $\hat{\mathbf{n_{1}}}$ and $A_{\perp}$ is the component perpendicular to it.

With
\begin{align}
\begin{split}
    \widetilde{H}_{s}^{(1)} & =\mu \mathbf{S} \cdot \mathbf{B}+D\left[S_{\|}^{2}-\frac{S(S+1)}{3}\right]  \\
\end{split}
\end{align}
The matrix representations of $S_1 = \mathbf{S}\cdot \mathbf{\hat{n_1}}$, $S_2 = \mathbf{S}\cdot \mathbf{\hat{n_2}}$, and $S_3 = \mathbf{S}\cdot \mathbf{\hat{n_3}}$:-
\begin{align}
\begin{split}
    & S_{1}=\left(\begin{array}{lll}
1 & 0 & 0 \\
0 & 0 &  0 \\
0 & 0 & -1
\end{array}\right)  \\
\end{split}
\end{align}
\begin{align}
\begin{split}
     & S_{2}=\frac{i}{\sqrt{2} }\left(\begin{array}{ccc}
0 & -1 & 0 \\
1 & 0 & -1 \\
0 & 1 & 0
\end{array}\right) \\
\end{split} 
\end{align}
\begin{align}
    & S_{3}=\frac{-1}{\sqrt{2}}\left(\begin{array}{ccc}
0 & 1 & 0 \\
1 & 0 & 1 \\
0 & 1 & 0
\end{array}\right)
\end{align}

\begin{align}
\widetilde{H}_{s}^{(1)} & =\left(\begin{array}{ccc}
\mu B_{1} & \mu \frac{-B_3 - iB_2}{\sqrt{2}}  & 0 \\
\mu \frac{-B_3 + iB_2}{\sqrt{2}} & -D & \mu \frac{-B_3 - iB_2}{\sqrt{2}} \\
0 &\mu \frac{-B_3 + iB_2}{\sqrt{2}}& -\mu B_{1}
\end{array}\right)+\frac{D}{3} I 
\end{align}
and
\begin{align}
\widetilde{H}_{s}^{(2)} & =-\hbar \mathbf{S} \cdot \frac{\mathbf{L}}{\mathbf{I}} \\
& =-\left(\begin{array}{ccc}
\hbar \frac{L_{1}}{I} & \frac{\hbar}{I} \frac{-L_{3}-i L_{2}}{\sqrt{2}} & 0 \\
\frac{\hbar}{I} \frac{-L_{3}+i L_{2}}{\sqrt{2}} & 0 & \frac{\hbar}{I} \frac{-L_{3}-i L_{2}}{\sqrt{2}} \\
0 & \frac{\hbar}{I} \frac{-L_{3}+i L_{2}}{\sqrt{2}} & -\hbar \frac{L_{1}}{I}
\end{array}\right) . 
\end{align}

\begin{align}
   \widetilde{H}_{s}^{(2)} =-\left(\begin{array}{ccc}
\hbar \omega_{1} & \hbar \frac{-\omega_{3}-i \omega_{2}}{\sqrt{2}} & 0 \\
\hbar \frac{-\omega_{3}+i \omega_{2}}{\sqrt{2}} & 0 & \hbar \frac{-\omega_{3}-i \omega_{2}}{\sqrt{2}} \\
0 & \hbar \frac{-\omega_{3}+i \omega_{2}}{\sqrt{2}} & -\hbar \omega_{1}
\end{array}\right)
\end{align}
where the effective Hamiltonian $\widetilde{H}_{s} $ is:-
\begin{equation}
\widetilde{H}_{s}=\widetilde{H}_{s}^{(1)}+\widetilde{H}_{s}^{(2)} 
\end{equation}

With the initial conditions, $\dot{\gamma}(0)=\omega_{0}$ and $\dot{\alpha}(0)=\dot{\beta}(0)=0$, and for the initial angles $\beta(0)=\beta_{0}$ and $\alpha(0)= \pi /2 ; \gamma(0)=0$; $\alpha(t) = \pi /2$, $\beta(t) = \beta_0$ and $\gamma(t) = \omega_0 t$

\begin{equation}
     \boldsymbol{\omega}=\left[\begin{array}{c}
\omega_{1}\\
\omega_{2} \\
\omega_{3}
\end{array}\right]=\left[\begin{array}{c}
\omega_0 \\
0 \\
0
\end{array}\right],
 \end{equation}

\begin{align}
\begin{split}
\vec{B}_{\text{body}} =
\begin{bmatrix}
B_1\\
B_2 \\
B_3
\end{bmatrix} = \begin{bmatrix}
B_x\cos\beta_0 + B_y \sin \beta_0  \\
B_x\sin\beta_0 \sin\gamma  - B_y \cos \beta_0 \sin\gamma \\
-B_x\sin\beta_0 \cos\gamma  +B_y\cos \beta_0\cos\gamma
\end{bmatrix}
\end{split}
\end{align}
Taking $B_{||} = (B_x\cos\beta_0 + B_y \sin \beta_0)$ and $|B_{\perp}| = (B_x \sin \beta_0 - B_y \cos \beta_0)$, with $B_{\perp} \sin \gamma = B_2$ and $B_{\perp} \cos \gamma = B_3$:-
\begin{align}
\widetilde{H}_{s}^{(1)} & =\left(\begin{array}{ccc}
\mu B_{\|} & \mu \frac{B_{\perp}}{\sqrt{2}} e^{-i\omega_0 t} & 0 \\
\mu \frac{B_{\perp}}{\sqrt{2}} e^{i\omega_0 t} & -D & \mu \frac{B_{\perp}}{\sqrt{2}} e^{-i\omega_0 t} \\
0 & \mu \frac{B_{\perp}}{\sqrt{2}} e^{i\omega_0 t} & -\mu B_{\|} 
\end{array}\right)+\frac{D}{3} I 
\end{align}

 \begin{align}
   \widetilde{H}_{s}^{(2)} =-\left(\begin{array}{ccc}
\hbar \omega_{0} & 0 & 0 \\
0 & 0 & 0 \\
0 & 0 & -\hbar \omega_{0}
\end{array}\right)
\end{align}

    \begin{align}
\widetilde{H}_{s} & =\left(\begin{array}{ccc}
\mu B_{\|} - \hbar \omega_0& \mu \frac{B_\perp}{\sqrt{2}}e^{-i \omega_0 t}  & 0 \\
\mu \frac{B_\perp}{\sqrt{2}}e^{i \omega_0 t} & -D &\mu \frac{B_\perp}{\sqrt{2}}e^{-i \omega_0 t}\\
0 &\mu \frac{B_\perp}{\sqrt{2}}e^{i \omega_0 t}& -\mu B_{\|}+ \hbar \omega_0
\end{array}\right)+\frac{D}{3} I 
\end{align}

Using the Feshbach formalism \cite{Feshbach1958,Feshbach1962,Band2022}  we have
\begin{align}
    H^{\text{eff}}_{NV} = \ &
    \begin{pmatrix}
        \mu B_{\|} - \hbar \omega_0 &
        \dfrac{\mu^2 B_{\perp}^2 e^{-2i\gamma}}{2D}\\[3pt]
        \dfrac{\mu^2 B_{\perp}^2 e^{2i\gamma}}{2D} &
        -\mu B_{\|} + \hbar \omega_0
    \end{pmatrix} \nonumber \\
    &+\left(\frac{D}{3}+\frac{\mu^2 B_{\perp}^2}{2D}\right) I 
\end{align}

\section{Maximum superposition size}
\label{appendix:superposition_size}
The equation of motion of Eqs. \eqref{eq:x_position2} and \eqref{eq:y_position2} can be solved analytically, considering the small angle of $\beta(t) \rightarrow \beta_0$, we have
\begin{align}
    x(t) & = \left(\frac{s \mu \eta}{m \Omega^2} -\frac{B_0}{\eta}\right)[\cos{(\Omega t)-1}], \\
    y(t) & = y_0\cos{(\Omega t)}-\frac{s \mu \eta}{m \Omega^2}\beta_0 [\cos{(\Omega t)-1}],
\end{align}
The maximum superposition in the two directions $x$ and $y$ occurs at $t_{max}=\pi/\Omega$ or half of the closed time of the SGI, $t_{closed}$, substituted the value of $s=\pm1$ as both arms of the SGI, and we have
\begin{align}
    \abs{\Delta x_{max}} & = \frac{4 \mu \eta}{m \Omega^2}, \\
    \abs{\Delta y_{max}} & = \frac{4 \mu \eta}{m \Omega^2}\beta_0,
\end{align}
and the combination of two directions is
\begin{align}
    {\Delta r_{max}} & = \frac{4 \mu \eta}{m \Omega^2}\sqrt{1+\beta_0^2},
\end{align}

\section{Spin contrast from librational dynamics}
\label{appendix:spin contrast fromlibration}
We start from the total Hamiltonian
\begin{equation}
H_t = \frac{p_\beta^2}{2I} + \frac{p_\gamma^2}{2 I_3} 
+ \frac{(p_\alpha - p_\gamma \cos \beta)^2}{2 I \sin^2 \beta} 
+ s\mu (B_x \cos \beta + B_y \sin \beta).
\end{equation}
For $\beta \ll 1$, we use $\sin\beta \approx \beta$ and $\cos\beta \approx 1$.

Defining shifted operators around the instantaneous equilibrium angle,
\begin{align}
\hat{\beta}' &= \hat{\beta} - \bar{\beta}(t), \\
\hat{p}_\alpha' &= \hat{p}_\alpha - \langle \hat{p}_\alpha \rangle, \\
\hat{p}_\gamma' &= \hat{p}_\gamma - \langle \hat{p}_\gamma \rangle,
\end{align}
with
\begin{equation}
\bar{\beta}(t) \approx \beta_0 + 
\frac{s \mu \big(B_x(t)\beta_0 - B_y(t)\big)}{I \omega_0^2},
\end{equation}
the Hamiltonian linearized in $\beta'=\beta-\bar{\beta}$ reads
\begin{equation}
H_t \approx 
\frac{p_\beta^2}{2I} 
+\frac{I\omega_0^2}{2}\beta'^2 
- f(p_\alpha',p_\gamma')\,\beta' 
+ g(p_\alpha',p_\gamma',t),
\end{equation}
where
\begin{equation}
f = 
\frac{(p_\alpha' - p_\gamma')^2}{I \beta_0^3} 
+ \frac{ \omega_0}{\beta_0}(p_\alpha' - p_\gamma'),
\end{equation}
and
\begin{equation}
g = 
\frac{(p_\alpha' - p_\gamma')^2}{2 I \beta_0^2}
- \omega_0 p_\gamma' 
- 
\frac{s \mu \big(B_x(t)\beta_0 - B_y(t)\big)}{I \omega_0} 
(p_\alpha' - p_\gamma').
\end{equation}

The initial state is taken as
\begin{widetext}
    \begin{align}
|\Psi(0)\rangle = 
\int \frac{dp_\alpha' dp_\gamma'}{\sqrt{2\pi \Delta p_\alpha \Delta p_\gamma}} 
\exp\!\left[-\frac{p_\alpha'^2}{4\Delta p_\alpha^2}\right]
\exp\!\left[-\frac{p_\gamma'^2}{4\Delta p_\gamma^2}\right]
|p_\alpha',p_\gamma'\rangle 
\otimes |0\rangle_{\bar{\beta}_0}.
\end{align}
\end{widetext}

The equilibrium position including spin and momentum shifts is
\begin{equation}
\bar{\beta}(s,t,p_\alpha',p_\gamma') = 
\beta_0 
+ \frac{s \mu (B_x(t)\beta_0 - B_y(t))}{I \omega_0^2}
+ \frac{f(p_\alpha',p_\gamma')}{I \omega_0^2}.
\end{equation}

The time-evolved state is
\begin{widetext}
    \begin{align}
|\Psi(t)\rangle
=
\int \frac{dp_\alpha' dp_\gamma'}{\sqrt{2\pi\Delta p_\alpha\Delta p_\gamma}}
e^{-\frac{p_\alpha'^2}{4\Delta p_\alpha^2}}
e^{-\frac{p_\gamma'^2}{4\Delta p_\gamma^2}}
e^{-\frac{i}{\hbar}\int^t g(p_\alpha',p_\gamma',t')dt'}
|p_\alpha',p_\gamma'\rangle\otimes |\kappa(t)\rangle,
\end{align}
\end{widetext}

with coherent state
\begin{equation}
|\kappa(t)\rangle
=
\left| -\sqrt{\frac{I\omega_0}{2\hbar}}\; 
\frac{s\mu(B_x(t)\beta_0 - B_y(t))}{I\omega_0^2}\;
e^{-i\omega_0 t} \right\rangle.
\end{equation}

Defining
\begin{equation}
A_\beta(t)=\frac{s\mu(B_x(t)\beta_0 - B_y(t))}{I\omega_0^2},
\end{equation}
we have
\begin{equation}
|\kappa(0)| = \sqrt{\frac{I\omega_0}{2\hbar}}\,A_\beta(0)
= \sqrt{\frac{I\omega_0}{2\hbar}}\,
\frac{s\mu (B_0\beta_0 + \eta y_0)}{I\omega_0^2},
\end{equation}
and
\begin{equation}
|\kappa(t_{\text{closed}})| = |\kappa(0)|.
\end{equation}

The spin contrast is defined as
\begin{equation}
C = \left| \langle \Psi_L(t_{\text{closed}}) | \Psi_R(t_{\text{closed}}) \rangle \right|.
\end{equation}
Using $\langle p'|p''\rangle\simeq \delta(p'-p'')$, one finds
\begin{align}
C =
\int \frac{dp_\alpha' dp_\gamma'}{2\pi \Delta p_\alpha \Delta p_\gamma}
e^{-\frac{p_\alpha'^2}{2\Delta p_\alpha^2}
      -\frac{p_\gamma'^2}{2\Delta p_\gamma^2}}
e^{-\frac{i}{\hbar}(\delta\alpha\,p_\alpha' + \delta\gamma\,p_\gamma')}
\left|\langle \kappa_L | \kappa_R\rangle\right|.
\end{align}

The phase difference is expressed as
\begin{equation}
\int_{0}^{t_{\text{closed}}}\!\! dt\,
\big[g_L - g_R\big]
=
\delta\alpha\, p_\alpha'
+\delta\gamma\, p_\gamma',
\quad \delta\alpha\simeq -\delta\gamma.
\end{equation}

The coherent-state overlap satisfies
\begin{align}
\left|\langle \kappa_L|\kappa_R\rangle\right|
&= \exp\!\left[-\frac{1}{2}\left|\kappa_L - (\kappa_R + \delta X)\right|^2\right],\nonumber\\
\delta X 
&= \sqrt{\frac{I\omega_0}{2\hbar}}\,
\frac{2\mu}{I\omega_0^2}
|B_0\beta_0 + \eta y_0|.
\end{align}

Using the bound
\begin{align}
\left|\langle \kappa_L|\kappa_R\rangle\right|
>
\exp\!\left[
-\frac{I\omega_0}{2\hbar}
\left(\frac{4\mu}{I\omega_0^2}|B_0\beta_0+\eta y_0|\right)^2
\right],
\end{align}
the final contrast becomes
\begin{align}
C &=
\exp\!\left[
-\frac{1}{2}
\left(
\delta \alpha^{2} \frac{\Delta p_{\alpha}^{2}}{\hbar^{2}}+
\delta \gamma^{2} \frac{\Delta p_{\gamma}^{2}}{\hbar^{2}}
\right)
\right]
\left|\langle \kappa_L|\kappa_R\rangle\right|\nonumber\\[4pt]
&>
\exp\!\left[
-\frac{1}{2}
\left(
\delta \alpha^{2} \frac{\Delta p_{\alpha}^{2}}{\hbar^{2}}+
\delta \gamma^{2} \frac{\Delta p_{\gamma}^{2}}{\hbar^{2}}+
\frac{8\mu^2(B_0\beta_0+\eta y_0)^2}{I\hbar\omega_0^3}
\right)
\right].
\end{align}

In the plot \ref{fig:contrast_mass} shows the relationship between contrast and the initial angular velocity. The contrast is described as the overlap between the two SGI arms wave packets, where a value of 1 represents perfect recombination, while a value of 0 means that the two wave packets do not overlap at all. In the calculation of the equations of motion, both spatial and rotational, an initial angular velocity of $\omega_0 = 2\pi \times 10$ kHz is used, for which the best contrast is obtained for a massive nanodiamond ($m = 10^{-16}$ kg, contrast $\sim 0.996$). The opposite trend is observed for small-mass nanodiamonds, so a higher initial rotational velocity is required to achieve a large contrast.

\section{Zero--Point Fluctuation Estimate for the Initial Displacement $y_0$}
\label{appendix:y_displacement}

The initial transverse displacement $y_0$ at the start of the interferometric sequence
can be estimated from the quantum fluctuations of a harmonic oscillator describing the
transverse center--of-mass motion.  
For a quantum harmonic oscillator with mass $m$, angular frequency $\omega$, and
excitation number $n$, the uncertainty of position rms is

\begin{equation}
    y_0^{(n)}=\sqrt{\frac{\hbar}{m\omega}\left(n+\frac12\right)} .
\end{equation}

For the parameters used in our simulations,
\[
m = 10^{-17}\,\mathrm{kg}, \qquad 
\omega = \sqrt{12.08} \simeq 3.476~\mathrm{rad/s},
\]
the prefactor evaluates to
\begin{equation}
    \frac{\hbar}{m\omega}
    = \frac{1.054\times10^{-34}}{(10^{-17})(3.476)}
    \approx 3.03\times10^{-18}\,\mathrm{m}^2 .
\end{equation}

Thus,
\begin{equation}
    y_0^{(n)} = \sqrt{3.03\times10^{-18}\,(n+0.5)} .
\end{equation}

Evaluating this expression for the ground state and for two excited states, $n=10$ and
$n=100$, we obtain
\begin{align}
    y_0^{(0)}   &\approx 1.23\times10^{-9}\,\mathrm{m}, \\
    y_0^{(10)}  &\approx 5.64\times10^{-9}\,\mathrm{m}, \\
    y_0^{(100)} &\approx 1.74\times10^{-8}\,\mathrm{m}.
\end{align}

These values represent the characteristic transverse displacement that the particle
possesses at the start of the interferometer due to quantum fluctuations alone.  
In the main text, we use $y_0$ chosen within this range depending on the assumed 
motional state of the nanodiamond.

\end{document}